\documentclass[]{interact}
\usepackage{subfigure}
\usepackage{amsmath}
\usepackage{graphicx}
\usepackage{dcolumn}
\usepackage{amsmath}
\usepackage{bm}
\usepackage{soul,color}
\usepackage{times}
\usepackage[numbers,sort&compress,merge]{natbib}

\theoremstyle{plain}

\theoremstyle{definition}

\theoremstyle{remark}

\begin{document}

\title{Quantum  memory and quantum correlations of Majorana qubits used for magnetometry}

\author{
\name{H. Rangani Jahromi$^{1}$\thanks{ Email: h.ranganijahromi@jahromu.ac.ir},  S. Haseli$^{2}$}
\affil{$^{1}$Physics Department, Faculty of Sciences, Jahrom
University, P.B. 74135111,  Jahrom, Iran.
}
\affil{
$^{2}$ Faculty of Physics,
Urmia University of Technology, Urmia, Iran.
}
}

\maketitle

\begin{abstract}
We address how the non-local nature of the topological qubits, realized by Majorana modes  and driven by an external magnetic field, can be used to control the non-Markovian dynamics of the system. It is also demonstrated that the non-local characteristic  plays a key role
 in control and protection of quantum correlations between Majorana qubits. Moreover, we discuss how those non-local qubits help us to enhance quantum magnetometry.

\end{abstract}

 \maketitle
\maketitle
\small
\section{Introduction\label{introduction}}

\par

Quantum correlations (e.g., entanglement \cite{Horodecki8652009} or discord \cite{Ollivier0179012002,Ferraro0523182010,Modi16552012}) are fundamental features of quantum mechanics and play  significant roles in various potential applications, such as  superdense coding, quantum teleportation, and quantum cryptography \cite{Bennett28811992,Ekert6611991,Datta0505022008,Pirandola12014}. Nevertheless, the quantum correlations are usually very fragile and broken by unwanted and  unexpected interactions with an environment referred to as quantum noise. In fact, the decoherence effect generates correlations between the system and environment, leading to an irreversible loss
of information from the system. Because there is no system which can be regarded as
truly isolated, investigating the the dynamics of the quantum correlations (inside the system) under
the action of noises and protecting  these quantum resources against the effects of the environment  are of great importance,
and also  major challenges for the realization of quantum computing devices \cite{Korotkov0401032010,Kim1172012,Man0123252012}.

Although the interaction with an environment causes the quantum
system to dissipate energy and lose its coherence \cite{Breuer2002}, this
process needs not be monotonic and the system
may recover temporarily  some of the lost  energy and/or
information. This behaviour called non-Markovianity
 can be characterized and quantified in different ways (see \cite{Rivas2014, Breuer2016} for reviews). In this paper, we use the measures which are easy to compute  and for which no external ancilla,  attached to  the open system of
 interest, is necessary to obtain the non-Markoianity.

 The  sudden change (death and birth) phenomenon (SCP) of quantum correlations and protecting them against the noise have
 been the themes of numerous works in the last few years. In the case of Markovian dynamics of quantum correlations, an
 interesting geometric interpretation of the sudden change  behavior of quantum correlations, for the simple situation in which
  Bell-diagonal states are considered, was provided in Ref. \cite{Lang1505012010}, while the conditions for the correlations to stay constant, the so called freezing phenomena, were investigated in Ref.
 \cite{You0121022012}, focusing on the phase damping channel. Moreover,  in Ref. \cite{Aaronson0121202013}, considering the case of Bell-diagonal states under the action
  of non-dissipative environments, the authors  proved that  all \textit{bona fide} measures of quantum correlations virtually present the freezing effect under the same dynamical
 conditions. Besides, using the quantum discord as a measure
  for quantum correlations, the authors of  \cite{Deb1853032015}  proved that a pure state  never present the SCP and that the freezing
 phenomena is not a general property of all the Bell-diagonal states.

 \par
 Because of memory effects, the non-Markovian channels may be more advantageous compared to   Markovian ones.
  The case of two independent qubits subject to two zero-temperature non-Markovian reservoirs was investigated in Ref. \cite{Wang0141012010}, in which the dynamics of entanglement and quantum discord were
 compared with each other. The authors verified that while the quantum discord can only vanish at some specific time periods, the entanglement  presents a sudden death such that it  disappears for all times after the
 critical time-point.
  In Ref. \cite{Fanchini0521072010}
 the authors discussed the similar problem, however they studied the case of a common
 reservoir, and again the SCP was  observed.  In addition, considering the
  class of Bell-diagonal states, the authors of  Ref. \cite{Xu3953042011} compared the non-Markovian dynamics of two geometric measures of quantum
 correlations  with  that of the quantum discord. Although all the three considered measures share a
 common sudden change point, one of the geometric measures does not present the
 freezing phenomenon. Besides, the
 dependence of the freezing effect on the choice of the correlation measure were also
 investigated in \cite{Yu04542013}. Moreover,  in Ref. \cite{Haikka0101032013}, the authors  studied the non-Markovianity and information flow for qubits subject to local dephasing with an Ohmic class
 spectrum and demonstrated  the existence of a temperature-dependent critical value of the Ohmicity parameter
 for the onset of non-Markovianity. They also unveiled a
 class of initial states for which the discord is forever frozen at a positive value. The investigation of the freezing phenomenon   is very important because it guarantees that the
 quantum protocols, in which quantum correlations are used as resources,  may be implemented such that they are unaffected by specific noisy conditions. Therefore, more analyses of
 the behaviour of quantum discord under different noisy quantum channels are necessary.

\par

 It has been seen that the topological quantum computation  is a promising scheme for realizing a quantum
 computer with robust qubits \cite{Nayak10832008}. In particular, there are new kinds of topologically ordered states, such as topological
 insulators and superconductors \cite{Fu0964072008,Hasan30452010,Qi10572011}, which are easy to realize physically. Among different  excitations for these these systems, the most interesting
 ones are the Majorana modes localized on topological defects,  obeying the non-Abelian
 anyonic statistics \cite{Wilczek6142009,Arovas7221984,Ivanov2682001}.   The Kitaev's
 1D spineless p-wave superconductor chain model \cite{Kitaev1312001,Sau0405022010,Alicea1253182010} is one of simplest scenario for realizing such Majorana modes. Each on-site fermion can be
 decomposed into two Majorana modes such that by
 appropriately tuning the model, the Majorana modes at the endpoints of Kitaev's chain may be
 dangling without pairing with the other nearby Majorana modes,  forming the usual fermions.
 Then, these two far separated endpoint Majorana modes can compose a topological qubit. The most important characteristic of the topological qubit is it is non-local, because the two Majorana modes are far separated. This non-locality causes
 the topological qubit to interact quite differently  with the environment comparing it to  the
 usual fermion. Motivated by this, we investigate how the non-local characteristic of the  topological qubits can be used for controlling their dynamics and probing the environment.
 
 Recent developments in the field of quantum metrology have shown that how quantum probes and quantum
 measurements allow us to achieve parameter estimation with precision beyond that achievable by any classical
 scheme \cite{Giovannetti2222011,Toth4240062014,Liu2019,RanganiAOP,RanganiQIP4,Rangani2018,RanganiAr1,RanganiQIP2019,RanganiSCR}. Estimating the strength of a magnetic field \cite{Albarelli1230112017,Ghirardi0121202018,Troiani2605032018,Danilin292018,RanganiIJQI2020} is a paradigmatic example in this respect, because it
 may be directly mapped to the problem of estimating the Larmor frequency for an atomic spin ensemble \cite{Wasilewsk1336012010,Koschorreck0936022010,Sewell2536052012,Ockeloen1430012013,Sheng1608022013,Lucivero1131082014,Muessel1030042014}.

\par In this paper the non-Markovian dynamics of a topological qubit realized by two Majorana modes coupled to a fermionic  Ohmic-like reservoir is  discussed in detail. The fermionic
environment is the helical Luttinger liquids realized
as interacting edge states of two-dimensional topological insulators. Imposing a cutoff for the linear spectrum of the edge states, we illustrate the significant role of this cutoff in controlling the non-Markovian evolution of the system and determining when the Ohmic-like environment can exhibit non-Markovian behaviour. Moreover, we study the non-Markovian behaviour of two independent Majorana qubits, each locally interacting with its own reservoir and investigate the effects of the imposed cutoff, originated from the non-local nature of the topological qubits, on quantum correlations and specially protection of them against the noise. In addition, we use the topological qubits   for the quantum magnetometry, i.e. quantum sensing of  magnetic fields by  quantum probes, and examine how their non-local characteristics help us to enhance the estimation.

 \par This paper is organized as follows: In Secion \ref{pre}, we
 present a brief review of the quantum correlation measures, quantum metrology  and non-Markovianity measures. The model is introduced in Section \ref{Model}. The non-Markovian behaviour of the single topological qubit is discussed in Sec. \ref{NMdynamicsq1} and the study is extended to two-qubit scenario in Sec. \ref{NMdynamics} in which the different measures of  quantum correlations are also addressed. Moreover, in Sec. \ref{magnetometry} 
 we explore the quantum magnetometry using entangled topological qubits. 
      Finally in Section \ref{conclusion}, the main results are summarized.

\section{The Preliminaries}\label{pre}
\subsection{Quantum Correlation measures \label{QC}}
\subsubsection{Concurrence}
\par In order to quantify the entanglement of the evolved state, we use  \textit{concurrence} \citep{Wootters WK,RanganiQIP2}.
 For $ X $-type structure states defined as 
  \begin{equation}\label{Xstate}
 \rho_X=\left(\begin{array}{cccc}
\rho_{11} &0&0&\rho_{14}  \\
 0&\rho_{22}&\rho_{23}&0  \\
  0&\rho_{32}&\rho_{33}&0 \\
 \rho_{41} &0&0&\rho_{44}  \\
 \end{array}\right),
 \end{equation}
 the concurrence may be obtained easily using a simple expression  given by \cite{WYSun}
\begin{equation}
C(\rho)=2\text{max}\left\lbrace 0,\Lambda_{1}(\rho),\Lambda_{2}(\rho) \right\rbrace , 
\end{equation}
in which
\begin{equation}
\Lambda_{1}(\rho)=|\rho_{14}|-\sqrt{\rho_{22}\rho_{33}}, \Lambda_{2}(\rho)=|\rho_{23}|-\sqrt{\rho_{11}\rho_{44}}.
\end{equation}
where $ \rho_{ij} $'s denote the elements of density matrix $ \rho^{X} $. The  concurrence is equal to zero for separable states and it equals to  $ 1 $ for maximally entangled states.

\subsubsection{Quantum Discord}
\par The usual quantum discord (QD), in terms of von Neumann
entropy, is  defined as difference between total correlations and classical correlations \cite{Lecture,Wang C-Z}:
\begin{equation}
QD(\rho_{AB})=I(\rho_{AB})-\mathcal{C}(\rho_{AB}).
\end{equation}
where
\begin{equation}
I(\rho_{AB})=S(\rho_{A})+S(\rho_{B})-S(\rho_{AB}).
\end{equation}
\begin{equation}
\mathcal{C}(\rho_{AB})=S(\rho_{A})-\text{min}_{{\left\lbrace\Pi^{B}_{k} \right\rbrace }}S(\rho_{A}|B),
\end{equation}
in which $ I(\rho_{AB}) $ denotes the quantum mutual information  measuring the total correlations, including
both classical and quantum, for a bipartite state $ \rho_{AB} $. Besides,   
$ S(\rho)=-\text{Tr}\left( \rho \text{log}_{2}\rho\right)   $ represents the von Neumann entropy of a quantum state, and $ \mathcal{C}(\rho_{AB}) $ is a measure of classical correlations.
The minimization is performed over all complete sets of projective measurements on subsystem $ B $. Moreover, $ S(\rho_{A|B})=\sum\limits_{k}^{}p_{k}S(\rho^{k}_{A}) $ represents  the conditional entropy
for subsystem A; $ p_{k}=\text{Tr}\big[(I_{A}\otimes \Pi^{B}_{k} )\rho_{AB}(I_{A}\otimes \Pi^{B}_{k} )\big] $ and $ \rho^{k}_{A}=\text{Tr}_{B}\big[(I_{A}\otimes \Pi^{B}_{k} )\rho_{AB}(I_{A}\otimes \Pi^{B}_{k} )\big]/p_{k} $ are respectively the probability and the state
of subsystem A obtaining  measurement outcome $ k $.

\subsubsection{Local quantum uncertainty}\label{LQUd}
\par
Although  the quantum correlations measures are usually
defined as a direct function of the density matrix $ \rho $ itself, its
other forms can also be advantageous. For example,  the square root $ \sqrt{\rho} $ in the well-known notion of the \textit{skew information} has been used to study the
local quantum uncertainty (LQU) which is an important measure of the quantum correlation.
Assuming  a bipartite quantum system prepared in quantum state $ \rho=\rho_{AB} $, we  suppose that $ O^{\varLambda}\equiv O^{\varLambda}_{A} \otimes \mathcal{I} _{B}  $ represents a local observable, in which $  O^{\varLambda}_{A}$ denotes a Hermitian operator on subsystem
$ A $ with non-degenerate spectrum $ \varLambda $.
The LQU with respect to  $ A $ is defined as follows \cite{Girolami2013,Ming2018}
          \begin{equation}\label{LQU}
\text{LQU}^{\varLambda}_{A}=\min_{O^{\varLambda}}I(\rho,O^{\varLambda}).
          \end{equation}
in which  $ I(\rho,O^{\varLambda})=-\dfrac{1}{2} \text{Tr}\{[\sqrt{\rho},O^{\varLambda}]^{2}\} $   represents the skew information. In addition, we should minimize over  all local observables of $ A $ with non-degenerate spectrum $ \varLambda $.

\subsubsection{Trace norm of discord}\label{TND}
For a bipartite system described by the density matrix $ \rho_{AB} $, the trace norm
of discord (TND),  a measure of the quantum correlation, is given by \cite{Ming2018}

          \begin{equation}\label{TNDD}
\mathcal{D}_{t}=\min_{\chi\in \mathcal{CQ}}||\rho_{AB}-\chi||,
          \end{equation}
where $ ||\rho_{AB}-\chi||=\text{Tr}\sqrt{(\rho_{AB}-\chi)^{\dagger}(\rho_{AB}-\chi)} $ represents the \textit{trace distance} between $ \rho_{AB} $
and $ \chi\in \mathcal{CQ}$ where  $ \mathcal{CQ}=\{\rho_{\mathcal{CQ}}\}$ is the set of classical-quantum states that can be written as

          \begin{equation}\label{CQ}
\rho_{\mathcal{CQ}}=\sum\limits_{k}^{}p_{k}\Pi^{A}_{k}\otimes \rho^{B}_{k},
          \end{equation}
a convex combination of the tensor products of the
orthogonal projectors $ \Pi^{A}_{k} $ for  subsystem A and  arbitrary density operators $ \rho^{B}_{k} $ for subsystem B, with $ \{p_{k}\} $ denoting any probability distribution. For a two-qubit X state $ \rho_{AB}\equiv \rho  $, the computation of the TND can be simplifed by \cite{Ming2018,Cheng2017,Ciccarello2014}

          \begin{equation}\label{TNdd}
\mathcal{D}_{t}=\frac{1}{2} \sqrt{\frac{\text{$\xi_{1} $}^2 \text{$\xi_{max} $}-\text{$\xi _{2}$}^2 \text{$\xi _{min}$}}{\text{$\xi _{1}$}^2-\text{$\xi _{2}$}^2+\text{$\xi _{max}$}-\text{$\xi _{min}$}}}
          \end{equation}
in which $ \xi_{1,2}=2(|\rho_{23}|\pm |\rho_{14}| ) $, $ \xi_{3}=1-2(\rho_{22}+\rho_{33}) $, $\xi_{max}=\max\{\xi^{2}_{3},\xi^{2}_{2}+x^{2}\}  $, and 
$\xi_{min}=\min\{\xi^{2}_{1},\xi^{2}_{3}\}  $ where $ x=2(\rho_{11}+\rho_{22})-1 $.

 \subsection{Non-Markovianity measures}\label{NMM}
 
 We first remind  some important definitions of the theory
 of open quantum systems. The time evolution of the density
 operator, describing the quantum state of an open  system,
 is
characterized by a time-dependent family of completely positive and
 trace preserving (CPTP) maps: $\mathcal{E}_{t}  $, called  the \textit{dynamical
 map}: $ \rho_{t}=\mathcal{E}_{t}(\rho_{0}) $, in which $ \rho_{0} $ denotes the density matrix of the open quantum system
 at  initial time $ t = 0 $.   Supposing that the inverse of $  \mathcal{E}_{t}$ exists for all times $ t\geq 0 $, and defining a two-parameter family of maps by
 means of 
 \begin{equation}
 \mathcal{E}_{t,t_{p}}\equiv \mathcal{E}_{t}\circ\mathcal{E}_{t_{p}}^{-1},
 \end{equation}
  one can write  CPTP map $ \mathcal{E}_{t} $   as a composite of the \textit{propagator} $ \mathcal{E}_{t,t_{p}} $ and $ \mathcal{E}_{t_{p}} $: 
 \begin{equation}\label{div}
\mathcal{E}_{t,0}=\mathcal{E}_{t,t_{p}}\circ\mathcal{E}_{t_{p},0}~,~~~~~\forall~~0 <t_{p}< t,
 \end{equation}
 where  $ \mathcal{E}_{t,0}\equiv\mathcal{E}_{t} $.  Although $ \mathcal{E}_{t,0} $ and
$ \mathcal{E}_{t_{p},0} $ should be completely positive by construction, the map $ \mathcal{E}_{t,t_{p}} $
 need not be completely positive and not even positive because
 the inverse $\mathcal{E}_{t_{p}}^{-1}  $ of a completely positive map $ \mathcal{E}_{t_{p}} $ need not be
 positive.
  Composition (\ref{div}), originating from  the existence of the inverse for all positive times, allows us to introduce the notion of  \textit{divisibility}. The family of dynamical maps is said to be P-divisible when propagator $ \mathcal{E}_{t,t_{p}} $ is positive as well as  trace-preserving, and CP-divisible if $ \mathcal{E}_{t,t_{p}} $ is
  CPTP for all $0 <t_{p}< t $ \cite{Breuer2016}. In the latter scenario, one can interpret $ \mathcal{E}_{t,t_{p}} $ as a legitimate
  quantum channel, mapping states at time $ t_{p} $ into states at
  time $ t $ \cite{Chakraborty0421052019}.
  
   In \cite{Rivas2010}, Rivas,
    Huelga, and Plenio (RHP) suggested that  the quantum evolution is called
  Markovian if and only if (iff) the corresponding dynamical
  map is CP-divisible. 
  Another  important characterization  of non-Markovianity was presented  by Breuer, Laine, and
  Piilo (BLP) who proposed that a non-Markovian
  process is characterized by a flow of information from the
  environment back into the open system \cite{Breuer2009,Laine2010}. Assuming that $\mathcal{E}_{t}  $ is invertible, one can show  that under this
  condition the quantum process is Markovian iff $\mathcal{E}_{t}  $ is P-divisible \cite{Wissmann2015,Breuer2016}.

 The most important common features of all non-Markovianity measures
 presented in the following subsections is that they are founded on the
 nonmonotonic time evolution of certain quantities when  the divisibility property is violated.
  Nevertheless, the inverse is not necessarily true; i.e., there may be
 nondivisible maps consistent with monotonic time evolution.
 
 \subsubsection{Breuer, Laine, Piilo (BLP) measure}\label{BLPmeasureS}
 The trace norm defined by $\parallel \rho \parallel=\text{Tr}\sqrt{\rho^{\dagger}\rho} =\sum\limits_{k}\sqrt{a_{k}} $, in which $ a_{k} $'s denote the eigenvalues of $\rho^{\dagger}\rho  $, yields  a 
 significant measure for the distance between two  states $ \rho^{1} $ and $ \rho^{2} $
 called \textit{trace distance} \cite{Lecture} $ D(\rho^{1},\rho^{2})=\frac{1}{2}\parallel \rho^{1}-\rho^{2} \parallel $.
  It may be proven that the  trace distance $  D(\rho^{1},\rho^{2}) $ can be interpreted as the distinguishability between   states $ \rho^{1} $ and $ \rho^{2} $. Besides, the trace distance is \textit{contractive} for any CPTP map $ \mathcal{E} $  \cite{Breuer2016}, i.e., $ D\big(\mathcal{E}(\rho^{1}),\mathcal{E}(\rho^{2})\big) \leq D(\rho^{1},\rho^{2})$, for  any two  quantum states $ \rho^{1,2} $. 
  Because any  dynamical map $\mathcal{E}_{t}  $ describing  the dynamics of an open quantum system is necessarily CPTP, the trace distance between
 the time-evolved quantum states can never be larger than the
 trace distance between the initial states. Therefore, the
 dynamics decreases the distinguishability of the quantum states in comparison with the initial preparation. It should be noted that, this 
 fact does \textit{ not} declare that $ D\big(\rho^{1}(t),\rho^{2}(t)\big) $ where $ \rho^{1,2}(t)\equiv\mathcal{E}_{t}(\rho^{1,2}(0)) $ is  a monotonically
 decreasing function with respect to time \cite{Breuer2012}. 
 \par
 
 According to BLP definition \cite{Laine2010,Breuer2009}, for a divisible process, distinguishability of two initial quantum states $ \rho^{1,2} $ diminishes continuously
 over time. Hence,
 a quantum evolution, mathematically defined by a quantum
 dynamical map $ \mathcal{E}_{t} $, is called Markovian when, for all pairs of initial quantum
  states $ \rho^{1}(0) $ and $ \rho^{2}(0) $,  trace
 distance $ D\big(\rho^{1}(t),\rho^{2}(t)\big) $
 decreases monotonically  at all instants.
  Hence, 
quantum Markovian dynamics  implies a continuous
 loss of information from the open system to the environment. On the other hand,  a non-Markovian  evolution is defined as a process in which, for certain time intervals,
 \begin{equation}
\sigma(t,\rho^{1,2}(0))\equiv \dfrac{d}{dt}D\big(\rho^{1}(t),\rho^{2}(t)\big) >0,
  \end{equation}
 i.e., the information flows back into the system temporarily, originating from appearance of quantum memory effects.

Following this, the measure of non-Markovianity $ \mathcal{N}_{BLP} $ can be defined as \cite{Laine2010}

 \begin{equation}\label{BLP}
\mathcal{N}_{BLP}(\mathcal{E}_{t})=\max_{\rho^{1,2}(0)}\int\limits_{\sigma>0}dt\sigma(t,\rho^{1,2}(0)).
  \end{equation}
  where the maximization is performed over all the possible pairs of initial quantum states
 and  the  integration is calculated over all time intervals 
  in which $ \sigma $ is positive.  It has been proved that \cite{Wissmann2012,Wissmann2015}, for
  any non-Markovian  quantum evolution of a \textit{single qubit},
  the maximum is achieved for a pair of
  pure orthogonal
  initial states corresponding to antipodal points on the Bloch sphere surface. Nevertheless, some challenges may be presented when one wishes
  to consider higher-dimensional systems of qubits \cite{Fanchini2013,Addis2013,LainePRL}. Moreover, 
although the non-Markovian evolution defined in this way are always
nondivisible, the converse is not necessarily true \cite{Chruscinski2011,Addis2014}.

\subsubsection{Lorenzo, Plastina, Paternostro (LPP) measure}\label{LPP}
 Another important proposed
 method to witness non-Markovianity was suggested by Lorenzo
 \textit{et al} in \cite{Lorenzo2013}. First, these authors expand   $ \rho $, quantum state of a N-level open system, in the
 basis $ \{G_{j}\}^{N^{2}-1}_{j=0} $, in which  the identity  $ G_{0}=I/\sqrt{N} $ and $  \{G_{j}\}^{N^{2}-1}_{j=1} $ are the
 Hermitian generators of SU(N) algebra \cite{Alicki1987},
 
 \begin{equation}\label{GB}
 \rho(t)=\mathcal{E}_t(\rho(0))=\sum\limits_{\alpha=0}^{N^{2}-1}\text{Tr}[\rho(t)G_{\alpha}]G_{\alpha}=\frac{I}{N}+\sum\limits_{\alpha=1}^{N^{2}-1}\text{Tr}[\rho(t)G_{\alpha}]G_{\alpha}=\sum\limits_{\alpha=0}^{N^{2}-1}r_{\alpha}G_{\alpha},
   \end{equation}
with 
$ \vec{r}^{~t}=(1/\sqrt{N},\vec{\textbf{r}}) ^{~t}$ in which $ \vec{\textbf{r}}=(r_{1},...,r_{N^{2}-1})^{~t} $ is called the generalized Bloch vector, 
and where $r_{\alpha}\equiv\text{Tr}[\rho(t)G_{\alpha}] $.
Then, it can be proven  that the action of the dynamical map
can be considered as an \textit{affine transformation} of the initial state Bloch vector  $\vec{\textbf{r}_{0}}$ \cite{Rivas2014},

 \begin{equation}\label{GT}
\rho(t)=\mathcal{E}_t(\rho(0))\longleftrightarrow \vec{\textbf{r}}=\textbf{M}(t)\vec{\textbf{r}_{0}}+\vec{\textbf{c}}(t)
   \end{equation}
where $[\textbf{M}(t)]_{ij} =\text{Tr}[G_{i}\mathcal{E}_{t}(G_{j})] $ and $ [\vec{\textbf{c}}(t)]_{i}= \text{Tr}[G_{i}\mathcal{E}_{t}(I)]/N$ for $ i,j>0 $. As discussed in \cite{Lorenzo2013},  the time-variation of  $  |\text{det}[\textbf{M}(t)]| $  describes the change in volume of the set of states accessible
through the evolution of the reduced state. Because the
Markovian evolution reduces (or leaves invariant) the volume of
accessible states, the following measure has been proposed \cite{Lorenzo2013} to quantify
the non-Markovianity of a quantum evolution,

\begin{equation}\label{LPPM}
\mathcal{N}_{LPP}(\mathcal{E}_{t})=\int\limits_{\partial_{t} |\text{det}[\textbf{M}(t)]|>0}\text{d}t \dfrac{\text{d}}{\text{d}t}|\text{det}[\textbf{M}(t)]|.
  \end{equation}

\subsubsection{Chanda and Bhattacharya (CB) measure}

Quantum coherence originating from  the superposition principle plays a key role in quantum mechanics such that it is a significant resource in quantum information theory. 
For a quantum state with the density matrix $ \rho $, the $ l_{1} $-norm measure of quantum coherence \cite{Baumgratz T} quantifying the coherence through the off diagonal elements of the density matrix in the reference basis, is given by

\begin{equation}\label{a7}
\mathcal{C}_{l_{1}}\left(\rho \right)=\sum_{i,j\atop i\ne j}|\rho_{ij}|.
\end{equation}

In the   basis $ \{|i\rangle\}_{i=1,...,d} $, all of the diagonal
density matrices, i.e.,
$
\varrho=\sum\limits_{i=1}^{d}c_{i}|i\rangle\langle i|
  $
 are called \textit{incoherent states}, while the states of the  form  
 $
 |\Psi_{d}\rangle=\frac{1}{\sqrt{d}}\sum\limits_{j=1}^{d}\text{e}^{i\phi_{j}}|j\rangle, ~~0<\phi_{j}<2\pi,
$
are recognized as \textit{maximally coherent states}.

\par
 For incoherent dynamics described by the \textit{incoherent operation} $ \Lambda_{t} $, a CPTP map which always maps any
 incoherent state to another incoherent one, the authors of \cite{Chanda2016} proposed the following  non-Markovianity measure based on $ l_{1} $-norm measure of quantum coherence
 $ \dfrac{dC_{l_{1}}(\rho(t))}{dt} $
  \begin{equation}\label{CB}
 \mathcal{N}_{CB}(\Lambda_{t})=\max_{\rho(0)\in \Im^{c}}\int\limits_{_{\dfrac{dC_{l_{1}}(\rho(t))}{dt} >0}}^{} ~\dfrac{dC_{l_{1}}(\rho(t))}{dt} dt,
   \end{equation}
 in which maximization should be performed over all the initial coherent states $ \rho(0) $ creating the set $\Im^{c}  $.
 Because in the case of large systems (i.e., large d), the above maximization procedure  may be complex
 to compute,  a nonoptimized  version of the measure can be used in  those cases, 

 \begin{equation}\label{simlifiedCB}
 \mathcal{N}^{l}_{CB}(\Lambda_{t})=\max_{\rho(0)\in \{|\Psi_{d}\rangle\langle \Psi_{d}|\}}\int\limits_{_{\dfrac{dC_{l_{1}}(\rho(t))}{dt} >0}}^{} ~\dfrac{dC_{l_{1}}(\rho(t))}{dt} dt,
   \end{equation}
   Generally, the maximization over all possible initial states involved
   in quantifying non-Markovianity is necessarily demanding.
   Nevertheless, starting with any chosen set of initial states, one can
   always obtain lower bounds to the measure of non-Markovianity,
  and hence achieve a qualitative assessment of the non-Markovian
   character of the evolution \cite{DHAR2015}.
 \section{The Model   \label{Model}}
 \begin{figure}[ht]
                         \includegraphics[width=12cm]{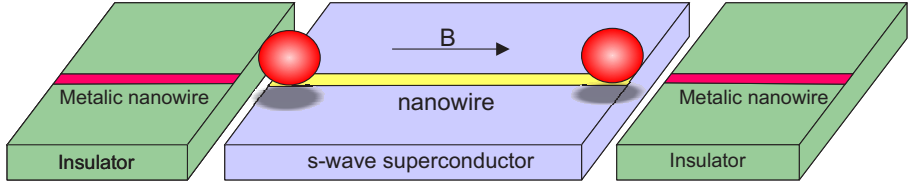}
                         \caption{\small Schematic diagram for the Majorana qubit coupled to the fermionic environment. }
                         \label{f11}
                           \end{figure}
 We consider a topological qubit realized by Majorana modes, generated at the endpoints of some nanowire with strong spin-orbit
 interaction, driven by an external magnetic field \textbf{B}  along the wire axis direction, and 
 placed on top of an s-wave superconductor (see Fig. \ref{f11}). Each of the Majorana modes  is
 coupled to the metallic nanowire via a tunnel junction  such that the  tunneling strength is controllable via an external gate voltage. 
 \par
  The total Hamiltonian is given by
  \begin{equation}
    H=H_{S}+H_{E}+V 
   \end{equation}
    where $ H_{s} $ is the Hamiltonian of the topological qubit  and $ V $ denotes the system-environment interaction Hamiltonian. Moreover, the  environment Hamiltonian is represented by $ H_{E} $ whose elementary
 constituents can be thought of as electrons. The decoherence affecting  the topological qubit  may be modelled as a fermionic
  Ohmic-like environment characterize by spectral density $ \rho_{spec}\propto \omega^{Q} $ with $ Q\geq 0 $.   The environment is called Ohmic for $ Q = 1 $, super-Ohmic for $ Q > 1 $ and sub-Ohmic for
  $ Q < 1 $.  This is realized by placing a metallic
  nanowire close to the Majorana endpoint. More specifically  the fermionic
  environment  chosen in this paper is the helical Luttinger liquids realized
  as interacting\textit{ edge states} of two-dimensional topological insulators \cite{Chao}. 
  Because these Majorana modes used for the qubits are zero-energy modes, we have $ H_{S}=0 $.
 Besides, the interaction Hamiltonian $ V $  is constructed by the electrons creation  (annihilation) operators, and Majorana modes $ \gamma_{1} $  
 as well as $ \gamma_{2} $ having the properties:
 
  \begin{equation}\label{MqubitProp}
\gamma^{\dagger}_{a}=\gamma_{a},~~~\{\gamma_{a},\gamma_{b}\}=2\delta_{ab},
    \end{equation}
 where $ a,b=1,2 $. The following representation can be chosen  for $ \gamma_{1,2} $:
 
   \begin{equation}\label{MajorPauli}
\gamma_{1}=\sigma_{1},~~~\gamma_{2}=\sigma_{2},~~~\text{i}\gamma_{1}\gamma_{2}=\sigma_{3},
     \end{equation}
 in which $ \sigma_{j} $'s denote the Pauli matrices.
 Before  turning on the interaction $ V $, the two
 Majorana modes form a topological (non-local) qubit with states $ |0\rangle $ and $ |1\rangle $ according to the following relations:
 
   \begin{equation}\label{Majorqubit}
 \frac{1}{2}(\gamma_{1}-\text{i}\gamma_{2})|0\rangle=|1\rangle,~~~~~\frac{1}{2}(\gamma_{1}+\text{i}\gamma_{2})|1\rangle=|0\rangle.
     \end{equation}
 
 Suppose that
 $ \varrho^{T} $, describing the state of the total system, is uncorrelated initially: $\varrho^{T}(0)=\varrho(0)\otimes \varrho_{E}$,
 where $\rho_{S}(0)$ and $\rho_{E}  $ denote the initial density matrices of the topological qubit and its environment, respectively. Assuming that the initial state of the Majorana qubit is given by
 
   \begin{equation}\label{InitialMajorqubit}
\varrho(0)=\left(\begin{array}{cc}
                   \varrho_{11}(0)&\varrho_{12}(0)  \\
                   \varrho_{21}(0)&\varrho_{22}(0)  \\
                  \end{array}\right),
      \end{equation}
 we can find that the reduced
 density matrix at time $ t $ is obtained as follows (for details, see \cite{HaoNJP}):

   \begin{equation}\label{Reduced1qMajorqubit}
\varrho(t)=\frac{1}{2}\left(\begin{array}{cc}
                   1+(2\varrho_{11}(0)-1)\alpha^{2}(t)&2\varrho_{12}(0)\alpha(t)  \\
                   2\varrho_{21}(0)\alpha(t)&1+(2\varrho_{22}(0)-1)\alpha^{2}(t) \\
                  \end{array}\right),
      \end{equation}
      where

   \begin{equation}\label{alpha}
\alpha=\text{e}^{-2B^{2}|\beta|I_{Q}},~~~~~\beta\equiv \dfrac{-4\pi}{\Gamma(Q+1)}(\dfrac{1}{\varGamma_{0}})^{Q+1}
      \end{equation}
      in which $ \varGamma_{0} $ represents   the high-frequency cutoff for the linear spectrum of the edge state and $ \Gamma (z) $ denotes the Gamma function. Moreover,

            \begin{equation}\label{IQ}
            I_{Q}= \left\{
            \begin{array}{rl}
            2\varGamma_{0}^{Q-1} \Gamma(\frac{Q-1}{2})\bigg[1-\,_1F_1\big(\frac{Q-1}{2};\frac{1}{2};-\frac{t^{2}\varGamma^{2}_{0}}{4}\big)\bigg]&~~~~~~~~ \text{for}~ Q\neq 1,\\
            \frac{1}{2}t^{2}\varGamma^{2}_{0} \,_2F_2\bigg(\{1,1\};\{3/2,2\};-\frac{t^{2}\varGamma^{2}_{0}}{4}\bigg)   & ~~~~~~~~\text{for}~ Q=1,
            \end{array} \right.
            \end{equation}
            where $ \,_pF_q $ denotes the \textit{generalized hypergeometric function} and $ \varGamma(z) $ represents  the Gamma function.
            
            \section{Non-Markovian dynamics of the one-qubit system}\label{NMdynamicsq1}
                       In this section,  the non-Markovian evolution of the qubit described by density matrix (\ref{Reduced1qMajorqubit}) is investigated. 
                       \begin{figure}[ht]
                        \includegraphics[width=12cm]{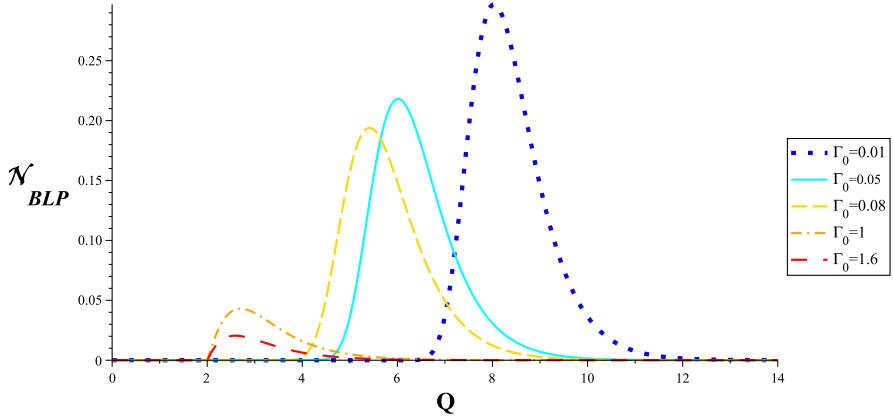}
                        \caption{\small Non-Markovianity $ \mathcal{N}_{BLP} $ versus $ Q $ for different values of $ \varGamma_{0} $. }
                        \label{f1}
                          \end{figure}
                       
                       In order  to compute non-Markovianity $ \mathcal{N}_{BLP} $, one has to find
                       a specific pair of optimal initial states maximizing the time
                       derivative of the trace distance. As pointed out in Sec. \ref{BLPmeasureS}, for
                       any non-Markovian evolution of a qubit, the maximal
                       backflow of information occurs for a pair of pure orthogonal
                       initial states corresponding to antipodal points on the surface
                       of the Bloch sphere. Using  numerical simulation, we can show
                       that for our model the optimal initial states are $ \{  |0\rangle, |1\rangle \}$. The trace
                       distance between these two states is $ D\big(\rho^{1}(t),\rho^{2}(t)\big)=\alpha^{2}(t) $, leading to $\mathcal{N}_{BLP}=\int\text{d}t \big(\dfrac{\text{d}\alpha^{2}(t)}{\text{d}t}\big)  $ where the integration is performed over  region $ \dfrac{\text{d}\alpha^{2}(t)}{\text{d}t}>0 $. The results are illustrated in Fig.
                       \ref{f1}: as apparent from the plot, there is a critical value of $ Q $ at which the non-Markovianity appears and an optimal value of $ Q $ for which the backflow of information from the environment to the system is maximized. 
                        Moreover, we see that an increase in cutoff $ \varGamma_{0} $ reduces the optimal value of $ Q $ at which $ \mathcal{N}_{BLP} $ is maximized and it also decreases the critical value of $ Q $ at which the non-Markovianity rises. The existence of a cutoff-dependent
                                    critical value of the Ohmicity parameter, ruling the Markovian
                                    to non-Markovian transition, is one of the main results of this
                                    paper.
                        Nevertheless, it is found that there exists a threshold value $ Q_{th}=2 $ such that when $ Q<2 $ the non-Markovianity completely vanishes, although increasing cutoff $ \varGamma_{0} $. 
                         By changing the $ Q $ parameter, we go from
                        sub-Ohmic reservoirs ($ Q < 1 $) to Ohmic ($ Q = 1 $) and super-Ohmic ($ Q > 1 $) reservoirs,
                        respectively.  Therefore, when the fermionic environment is Ohmic or sub-Ohmic the evolution of the Majorana qubit does not  exhibit non-Markovian effects. 
                          On the other hand, as  seen from the figure, the dynamics can be strongly non-Markovian for  small values of the cutoff, while an increase in cutoff suppresses the non-Markovianity and memory effects.
                   
                 Wee introduce parameter $ \kappa\equiv \big(K+\dfrac{1}{K}\big)/4 $, where $ K $  characterizes the Luttinger liquid, e.g.,
         $\kappa=1/2  $ for Fermi liquid. A rough estimation of the Luttinger parameter $ K $ is given by $ K^{2} \sim \big(1+\dfrac{U}{2\epsilon_{F}}\big)$
        in which $ \epsilon_{F} $ represents the Fermi energy and $ U\sim \dfrac{e^{2}}{\epsilon a_{0}} $, where $ \epsilon$ and  $ a_{0} $,  respectively, are the dielectric constant and
          the lattice length, denotes the characteristic Coulomb energy of the wire \cite{Kane12201992}. It is known that  the Coulomb interaction of the metallic wire can be tuned by choosing a different insulating substrate or gating.
           Therefore, the value
         of $ \kappa $ may be tuned by varying the effective repulsive/attractive short range interactions in the
         wire. Recalling  that $ Q=2\kappa-1 $, we should note that parameter $ \kappa $ characterizes the interaction strengths of the
         Luttinger/Fermi liquid. The larger  $ \kappa $, the stronger the correlation/interaction exhibited by the
         Luttinger liquid nanowire. Combining these points with the results obtained from Fig. \ref{f1}, we find that the non-Markovianity vanishes when the environments are so weakly ($ \kappa< 3/2 $) or  very strongly correlated ($ \kappa\rightarrow\infty $). Moreover, although it has been assumed that the coupling between the Majorana modes and the environment to be
                  weak such that the Gaussian approximation holds well  \cite{HaoNJP}, the non-Markovianity, controllable by the cutoff, may  occurs with correlated environments.
         \begin{figure}[ht!]
           \subfigure[]{\includegraphics[width=7cm]{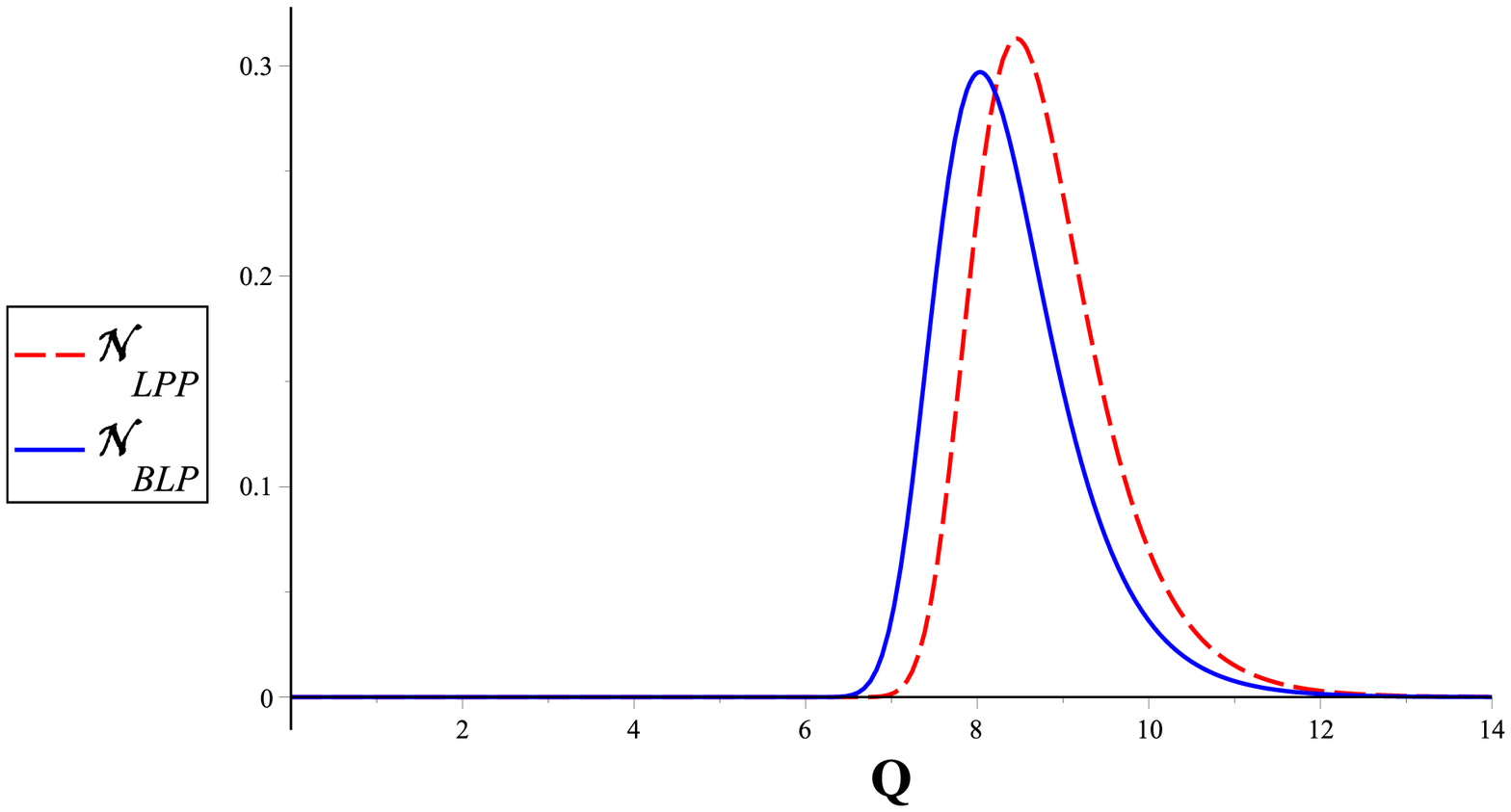}\label{N1BLPLPP} }
           \subfigure[]{\includegraphics[width=7cm]{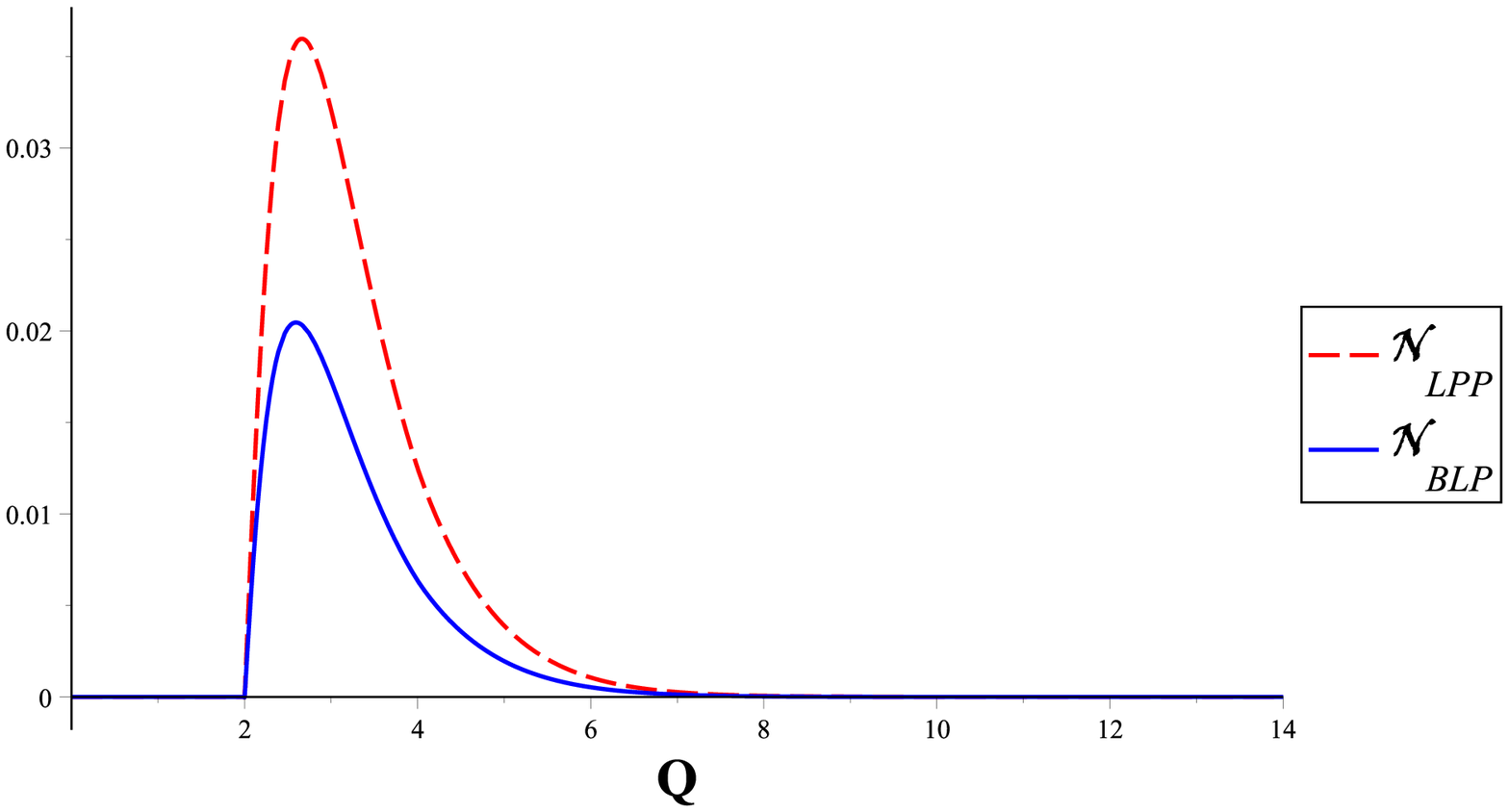}\label{N2BLPLPP} }
            \caption{The BLP and LPP measures of non-Markovianity versus Q for (a) $ \varGamma_{0}=0.01 $ and (b) $ \varGamma_{0}=1.6 $. }
              \end{figure}
         
         Now we compare BLP and LPP measures in the process of  detecting the non-Markovian dynamics of the topological qubit. Using the approach introduced in Sec. \ref{LPP}, one can show that $\textbf{M}(t)=\text{diag}(\alpha(t),\alpha(t),\alpha^{2}(t)) $, leading to $ \text{det}[\textbf{M}(t)]=\alpha^{4}(t) $ and consequently $\mathcal{N}_{LPP}=\int\text{d}t \big(\dfrac{\text{d}\alpha^{4}(t)}{\text{d}t}\big)  $ in which the integration should be performed over  region $ \dfrac{\text{d}\alpha^{4}(t)}{\text{d}t}>0 $. Plotting $ \mathcal{N}_{LPP} $, we again obtain exactly the similar results extracted from Fig. \ref{f1}. Nevertheless, as shown in Fig. \ref{N1BLPLPP} , the LPP measure is not as efficient as BLP measure in detecting the non-Markovianity when the cutoff is small. However, for larger values of the cutoff, the two measures qualitatively exhibit the same behaviour, i.e., approximately   the LPP is as sensitivity as BLP measure to detect non-Markovianity (see Fig. \ref{N2BLPLPP}).

            . 
            
             \section{Quantum correlations and non-Markovian dynamics of the two-qubit system}\label{NMdynamics}
            In order to study the dynamics of quantum correlations for Majorana qubits, we
            need the expression of the two-qubit density matrix. As known for the case that the subsystems interact independently with their environments, the complete dynamics of the two-qubit system
           can be computed by knowing the reduced density matrices of   each qubit \cite{Bellomoprl,Bellomopra} (see Appendix \ref{DMB}).
           We focus our analysis on the initial  entangled state $ \rho_{0}=|\psi_{0}\rangle \langle \psi_{0}| $ where 
           
            \begin{equation}\label{initialtwoqubit}
|\psi_{0}\rangle=\cos(\vartheta/2)|00\rangle+\sin(\vartheta/2)|11\rangle.
            \end{equation} 
           Therefore, the nonzero matrix elements of the
           evolved density matrix are given by

      \begin{equation}\label{MasterELEMENT}
                 \rho_{1,1}(t)=\dfrac{1}{4}\,{\alpha}^{4}+\dfrac{1}{2}\,\cos \left( \vartheta \right) {\alpha}^{2}+\dfrac{1}{4},
                 ~~
                 \rho_{2,2}(t)=\rho_{3,3}(t)=-\dfrac{1}{4}\,{\alpha}^{4}+\dfrac{1}{4},
                \nonumber 
                \end{equation}
                
                \begin{equation}
                \rho_{4,4}(t)=1-\bigg(\rho_{1,1}(t)+\rho_{2,2}(t)+\rho_{3,3}(t)\bigg);~~
          \rho_{1,4}(t)=\rho_{4,1}(t)= \frac{1}{2}\,{\alpha}^{2}\sin \left( \vartheta \right).
                 \end{equation}
               
      Computing (\ref{simlifiedCB}) for the above dynamics leads to a lower bound on the degree of non-Markovianity of the two-qubit  evolution.
     Because $ \mathcal{C}_{l_{1}}=\alpha^{2}(t)\text{sin}\vartheta $, it is found that this lower bound (with $ \vartheta=\pi/2 $) and the BLP measure for the one-qubit dynamics coincide exactly. Now  this phenomenon is discussed using the quantum correlations between the two qubits and their efficiency in detecting the non-Markovian behaviour of the composite system is investigated. According to our computed results,  the  bipartite entanglement, as quantified by concurrence, is given by
     
      \begin{equation}\label{Cuncurrence}
      C(\rho(t))=2\,\text{max} \left( 0,\dfrac{{\alpha}^{2}}{2}\sin \left( \vartheta \right) -
       \dfrac{\left| {\alpha}^{4}-1 \right| }{4} \right).           
        \end{equation}
     \begin{figure}[ht!]
                \subfigure[]{\includegraphics[width=7cm]{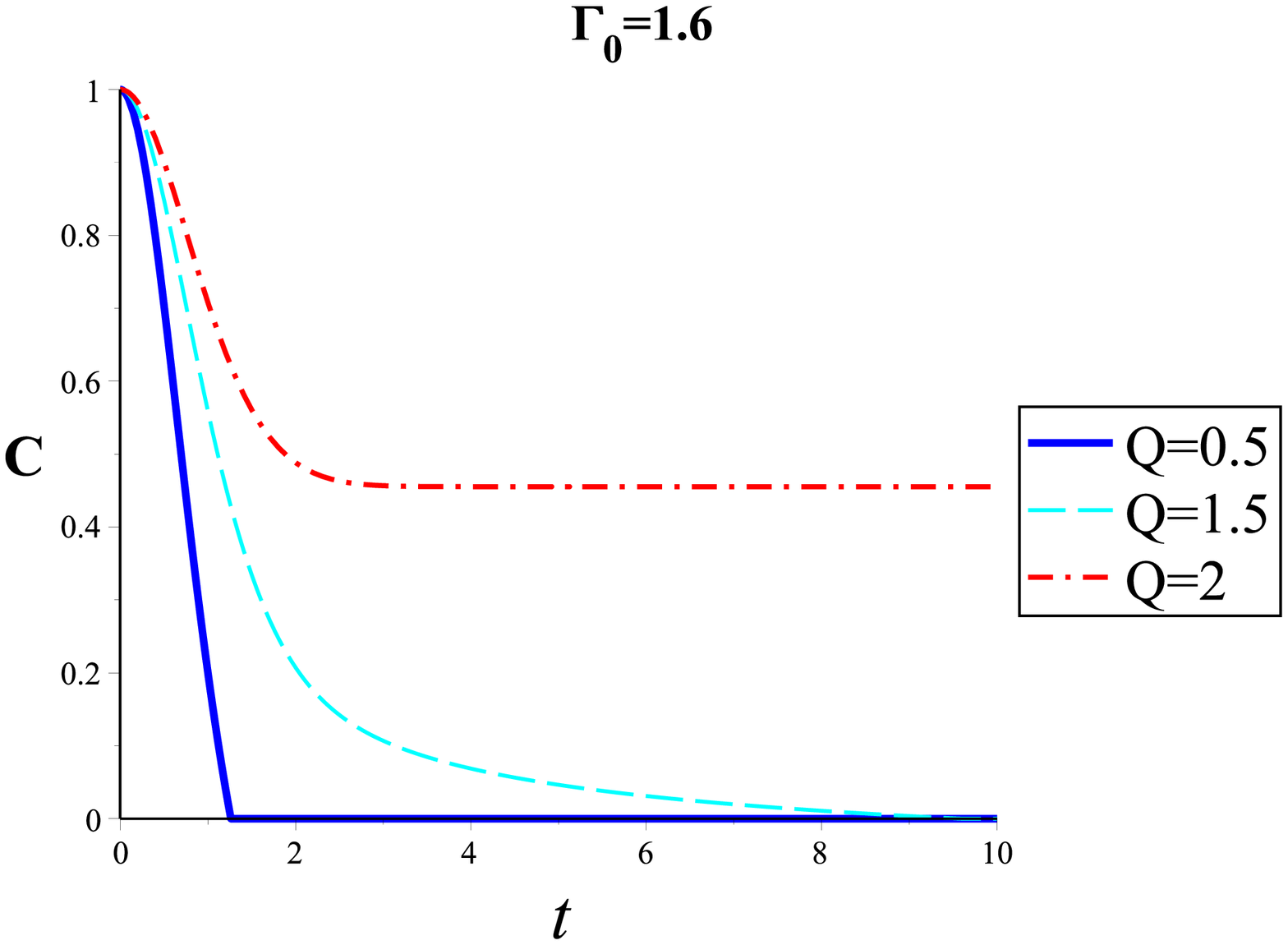}\label{CQQ} }
                \subfigure[]{\includegraphics[width=7cm]{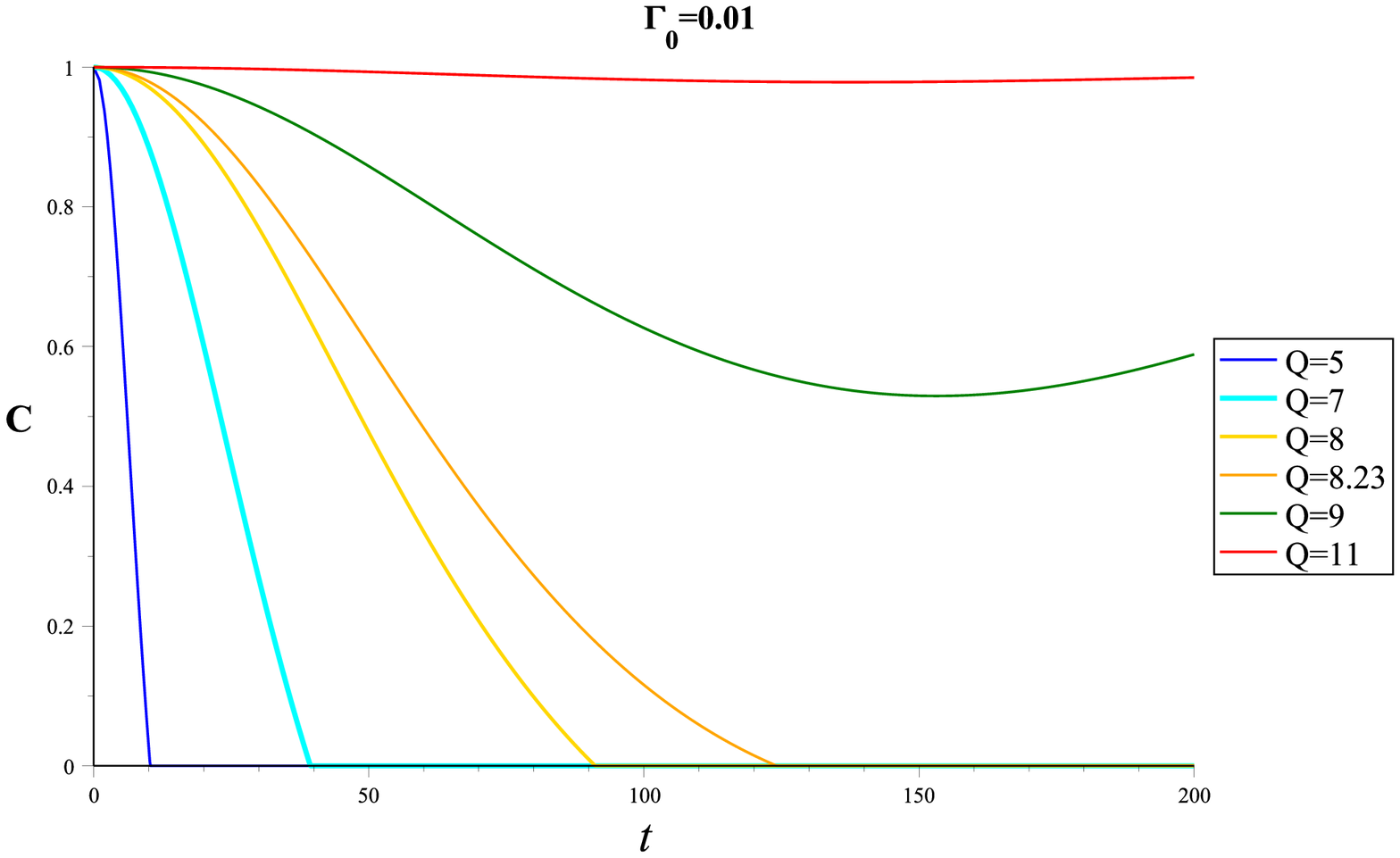}\label{CQ1} }
                \subfigure[]{\includegraphics[width=10cm]{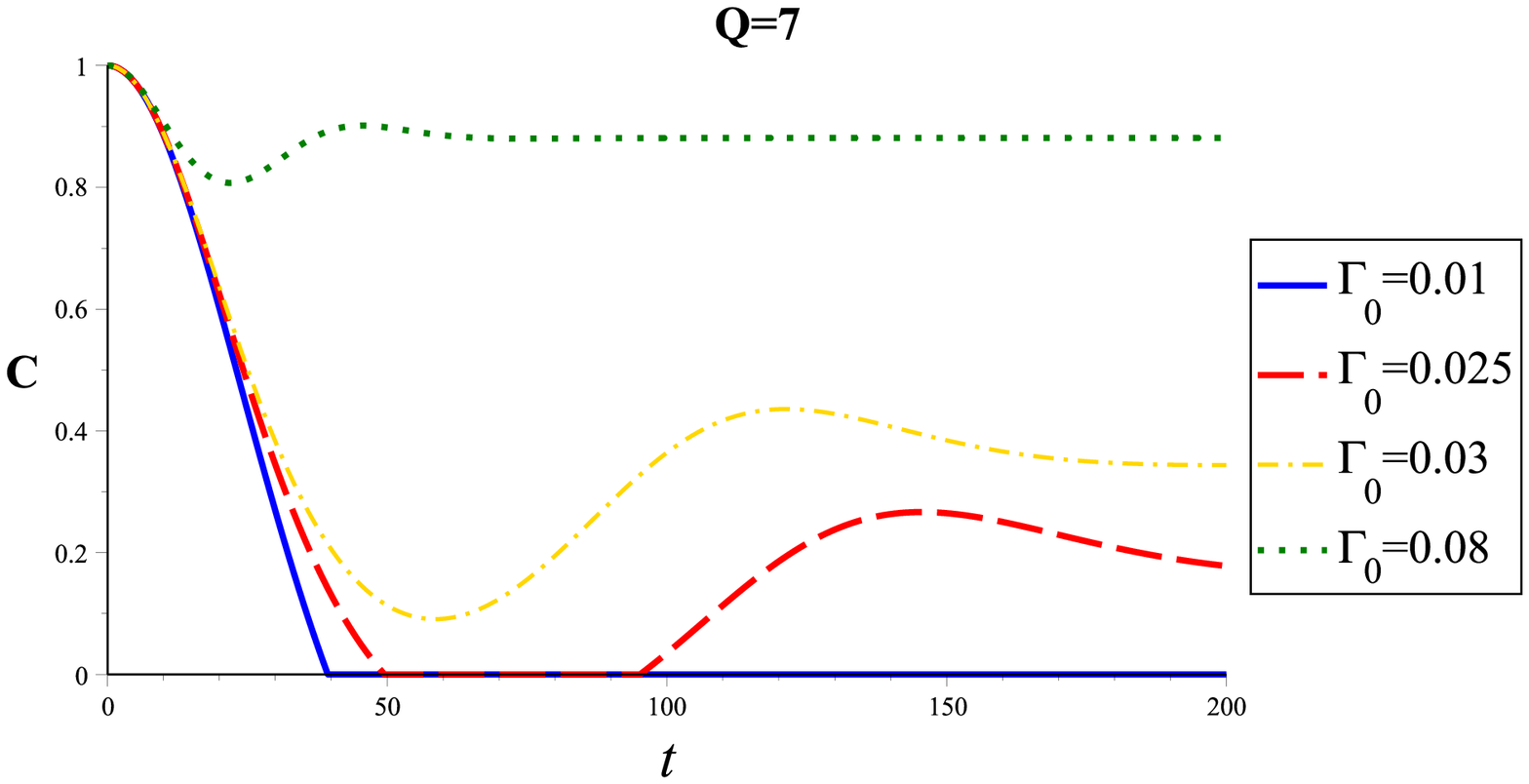}\label{CQ2} }
                 \caption{The time variation of the concurrence for (a), (b)   different values of Ohmicity parameter $ Q $ in Markovian and non-Markovian regimes. (c) The same quantity for different values of cutoff $ \varGamma_{0} $. }
                 \label{Concurrence} 
                   \end{figure}
          Figure \ref{Concurrence}   shows that the entanglement may decrease abruptly and non-smoothly to zero in a finite time due to the influence of
          quantum noise. This
          non-smooth finite-time decay is called entanglement sudden death (ESD) which   may occur in both Markovian and non-Markovian dynamics, as  seen in Fig. \ref{Concurrence}. In Markovian regime, if the  ESD occurs the entanglement cannot be restored over time, while  in non-Markovian regime  it is sometimes
          possible to recover the entanglement, however this recovery is not guaranteed. 
           In fact, in non-Markovian regime the  ESD can  occur only for small values of $ \varGamma_{0} $. The  ESD may exhibit robustness, even when the $ Q=Q_{\text{opt}} $ at which the non-Markovianity is maximized. However, increasing $ Q $ $ (Q>Q_{\text{opt}}) $ can remove the sudden death and  protect  the entanglement over time in both markovian and non-Markovian dynamics (see Figs. \ref{CQQ} and \ref{CQ1}). This revival phenomenon is
                     induced by the memory effects of the reservoirs, which
                     allows the two-qubit  to reappear their entanglement after a dark
                     period of time, during which the concurrence is zero.
           Another important feature illustrated in Fig. \ref{CQ2} is the high level of quantum control realized by imposing a cutoff $ \varGamma_{0} $. The plot shows that increasing the cutoff, we can remove the entanglement  ESD and also  protect  the initial maximal entanglement  over time. Moreover, we observe  the suppressive role of the cutoff on non-Markovian dynamics, leading to reduction of  the oscillations amplitude in entanglement dynamics.
           \begin{figure}[ht!]
                           \subfigure[]{\includegraphics[width=7cm]{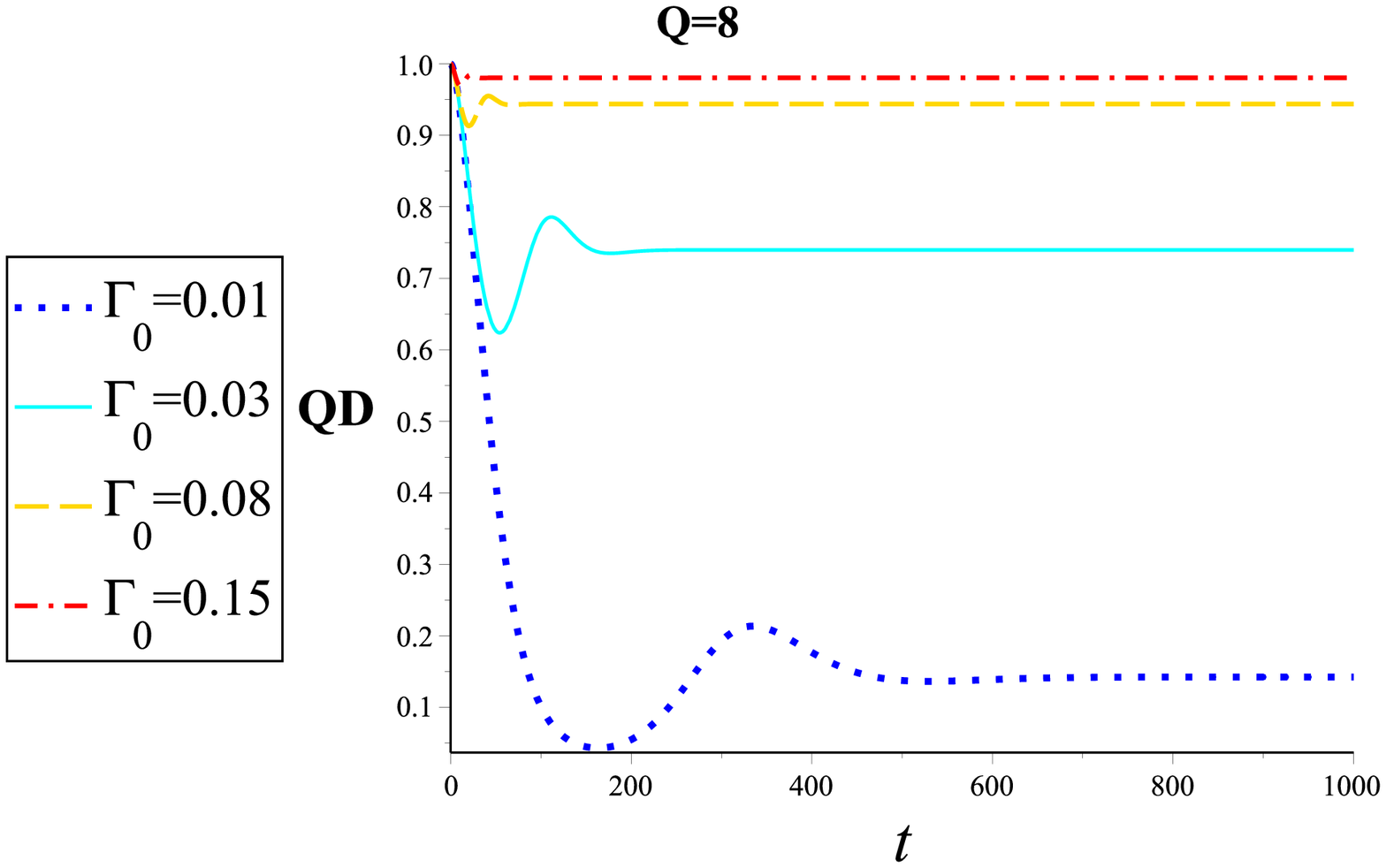}\label{QD1} }
                           \subfigure[]{\includegraphics[width=7cm]{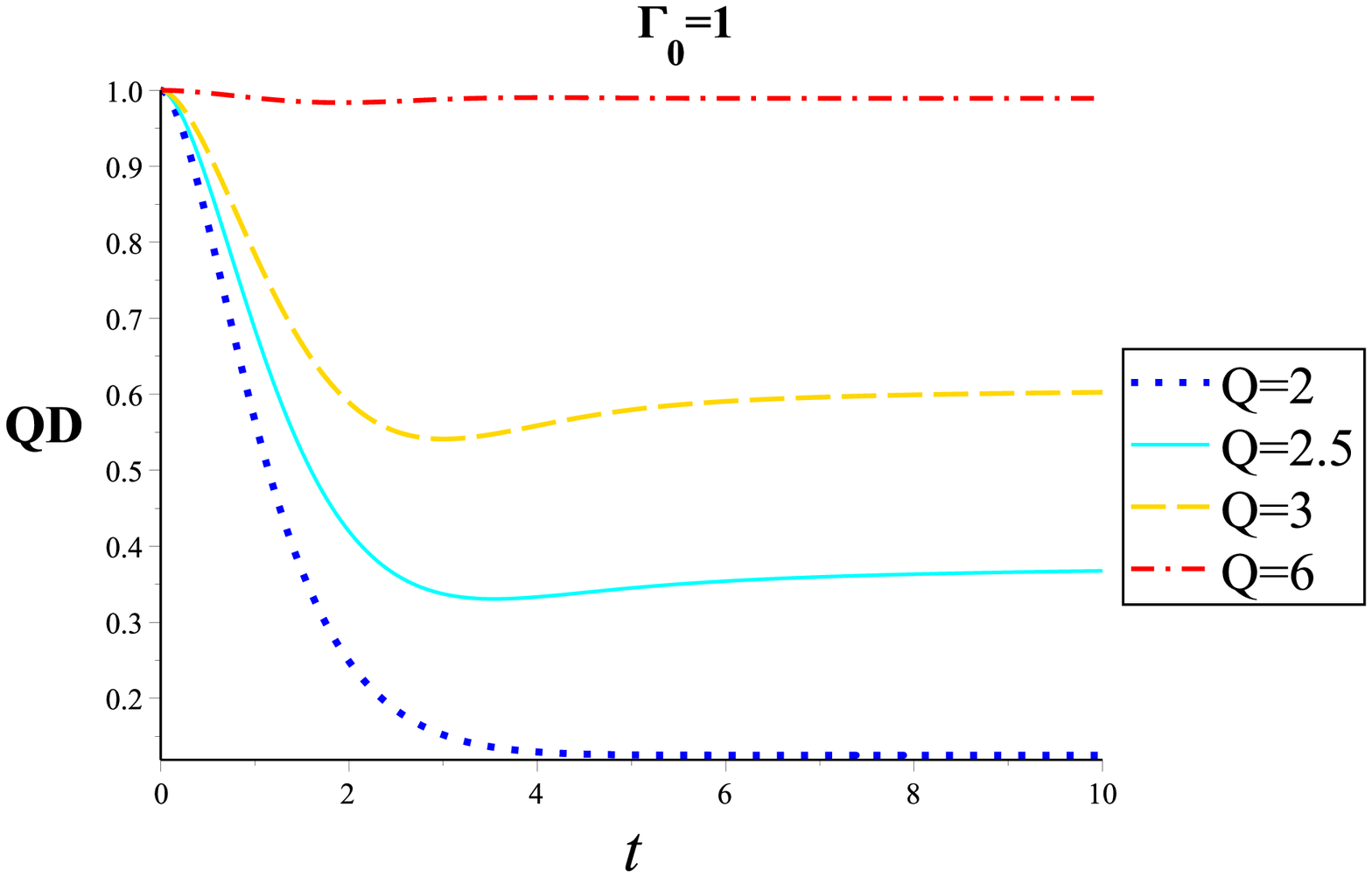}\label{QD2} }
                            \caption{(a) The QD as a function of time for   different values of cutoff $ \varGamma_{0} $. (b) The same quantity for different values of Ohmicity parameter $ Q $. }
                            \label{QuantumD} 
                              \end{figure}
                              \begin{figure}[ht]
                              \includegraphics[width=9cm]{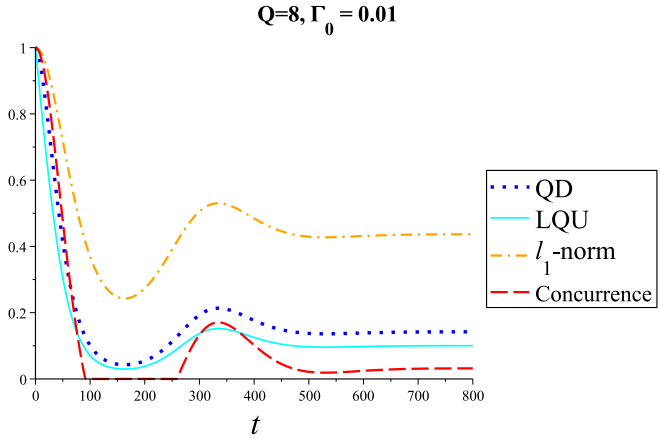}
                              \caption{\small Comparing the dynamics of  different quantum correlation measures with the coherence.}
                                           \label{diff}
                                                        \end{figure}
                              
        The analytical expression for the QD associated with the density matrix, whose elements are given in Eq. (\ref{MasterELEMENT}), is presented in Appendix \ref{QDLQU}. The dynamics of the QD is similar to one observed for the concurrence, except that the QD exhibits no sudden death (see Fig. \ref{QuantumD}). Another important characteristic of this measure of quantum correlations in our model  is that the time variation of the  QD   may be stopped and the QD can remain approximately constant over time after a while, a
        phenomenon known as \textit{freezing of QD}. In particular, Fig. (\ref{QuantumD}) clearly shows that increasing $ \varGamma_{0} $ or $ Q $ such that the non-Markovianity is suppressed, we can approximately obtain the \textit{time-invariant} discord. This phenomenon also may occur for $ Q<2 $ 
           for which the dynamics is always Markovian provided that a large cutoff is imposed. It is also interesting to compare the behaviour of different quantum correlation measures and coherence. We find that for initial state  (\ref{initialtwoqubit}), the TND and LQU, respectively, are given by
           
           \begin{equation}\label{AsliTND}
                TND =\frac{1}{2}\mathcal{C}_{l_{1}}=\frac{1}{2}\alpha^{2}(t)\text{sin}\vartheta ,~~~LQU_{\vartheta=\frac{\pi}{2}} =1-\sqrt{1-\alpha^{4}}.
                   \end{equation}
 A more general expression for the LQU is presented in Appendix \ref{QDLQU}. Figure \ref{diff} illustrates the dynamics of quantum correlations,  allowing us to compare them with the quantum coherence and hence  analyze their efficiency in detecting the non-Markovianity. We see that the quantum entanglement between the qubits is not a reliable witness to detect the non-Markovianity because there exists some period of time at which the quantum coherence increases while the entanglement has  vanished. However, the   QD or discord-like measures including LQU, and TND exhibit
 qualitatively the same dynamics as coherence. Therefore the revival of each of these quantum correlation measures can be attributed to the memory effects of the reservoirs. Moreover, we can conclude  that the coherence and quantum-discord like measures \textit{all} freeze under the same
 dynamical conditions,  albeit this phenomenon occurs after a while in the
 case of Markovian dynamics. 
           
        \par   
           We also observe that under the  dynamics considered here, discord-like measures are more robust than entanglement, and hence they are immune to the sudden death.
           	This important phenomenon  points to the fact that the absence of entanglement
           	does not necessarily lead to the absence of quantum correlations.  In addition, it suggests that quantum
           	computers based on the quantum correlations, different from those based on entanglement, are more robust against
           	the external perturbations and hence encourages us to try to  
           	implement an efficient quantum computer. Moreover,  it mentions why some
           	quantum algorithms can work well in the absence of
           	entanglement. In fact, the QD must be larger than zero in
           	the implementation of those algorithms in the situations that the
           	entanglement is missed or destroyed by decoherence.
           
       \par
   An important question which might be asked is why the ESD experienced by the concurrence, is not exhibited  by the discord-like measures. This phenomenon can be interpreted by analyzing the behavior of the coherence. As discussed by  some researchers, the QD is equal to the quantum coherence
   in a set of mutually unbiased bases for Bell-diagonal states \cite{Jin-Xing Hou0423242017}. Accordingly, the sudden birth and sudden death of
   the QD can be regarded as sudden birth and sudden death of quantum coherence in an optimal basis. In our work, because the quantum coherence  exhibits no sudden birth or death, the QD also shows the similar behavior. It should be noted that both the   quantum coherence and QD are  parts of total
   correlation, i.e., quantum mutual information, and coherence is a more basic quantum resource for quantum information tasks \cite{Jin-Xing Hou0423242017}.   The  difference between the QD and entanglement can also be  understood from a geometrical viewpoint:    sudden death of entanglement happens when the evolving density matrix crosses the boundary into the set of unentangled (separable)  states, where it can remain permanently, or for a finite period of time.  However, the QD vanishes only on a lower-dimensional manifold within the set of density matrices, and hence we would  expect the QD to  vanish completely at most only at particular instants of time, when the evolving density matrix crosses this manifold. More specifically, K. Modi et al., \cite{Modi0805012010}  proposed such geometrical interpretation of correlations which considers the concurrence 
   as the distance in Hilbert space between the quantum
   state of the system and the nearest separable
   state
   and  QD as the distance to the nearest classical pure state. From that perspective one may say that for most of
   the states in the family the separable states are close, while at the same time
   the classical pure states are far in Hilbert space. 
           
           \section{Quantum magnetometry \label{magnetometry}}
                    \begin{figure}[ht!]
                      \subfigure[]{\includegraphics[width=7cm]{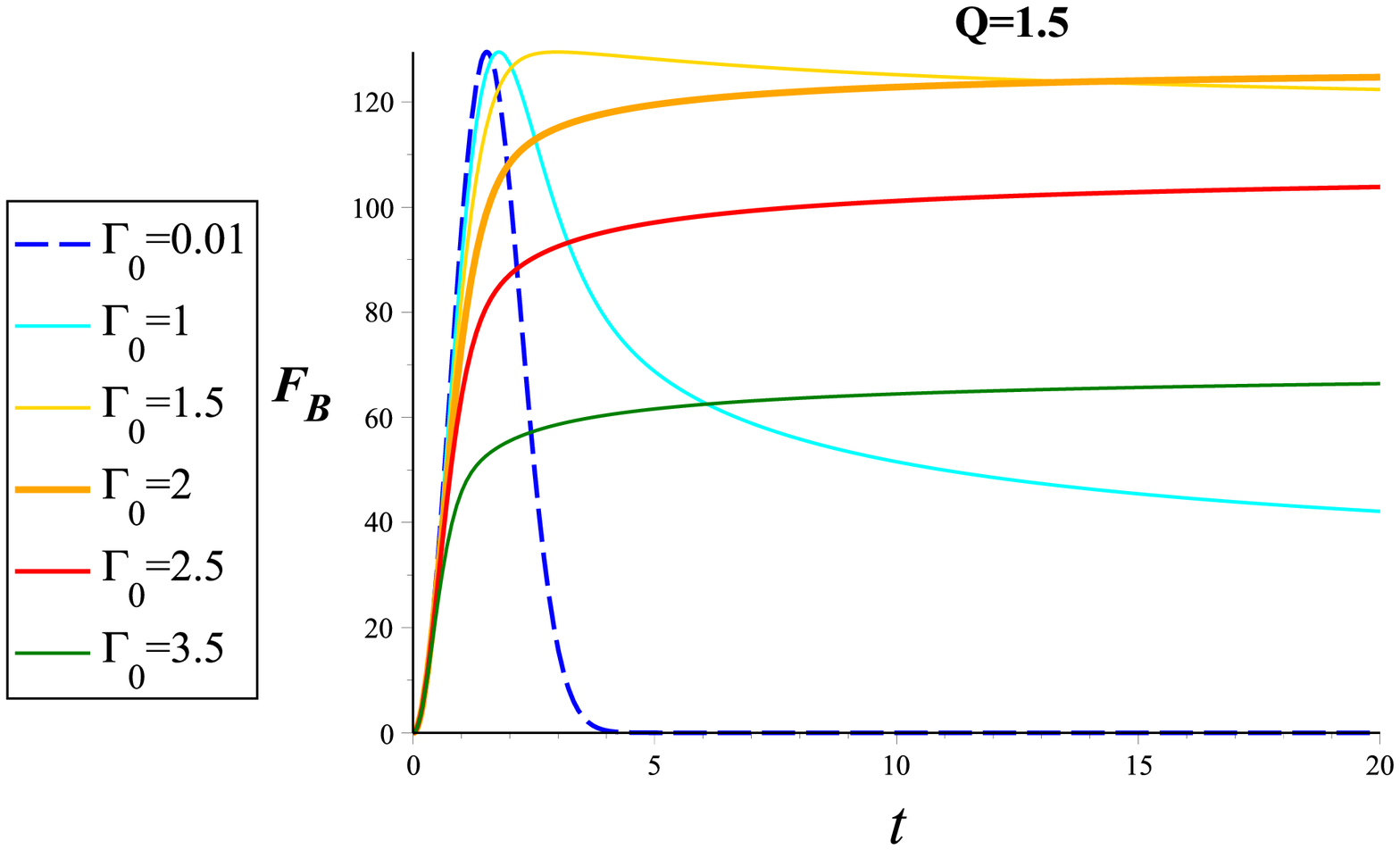}\label{QFIgama1} }
                      \subfigure[]{\includegraphics[width=7cm]{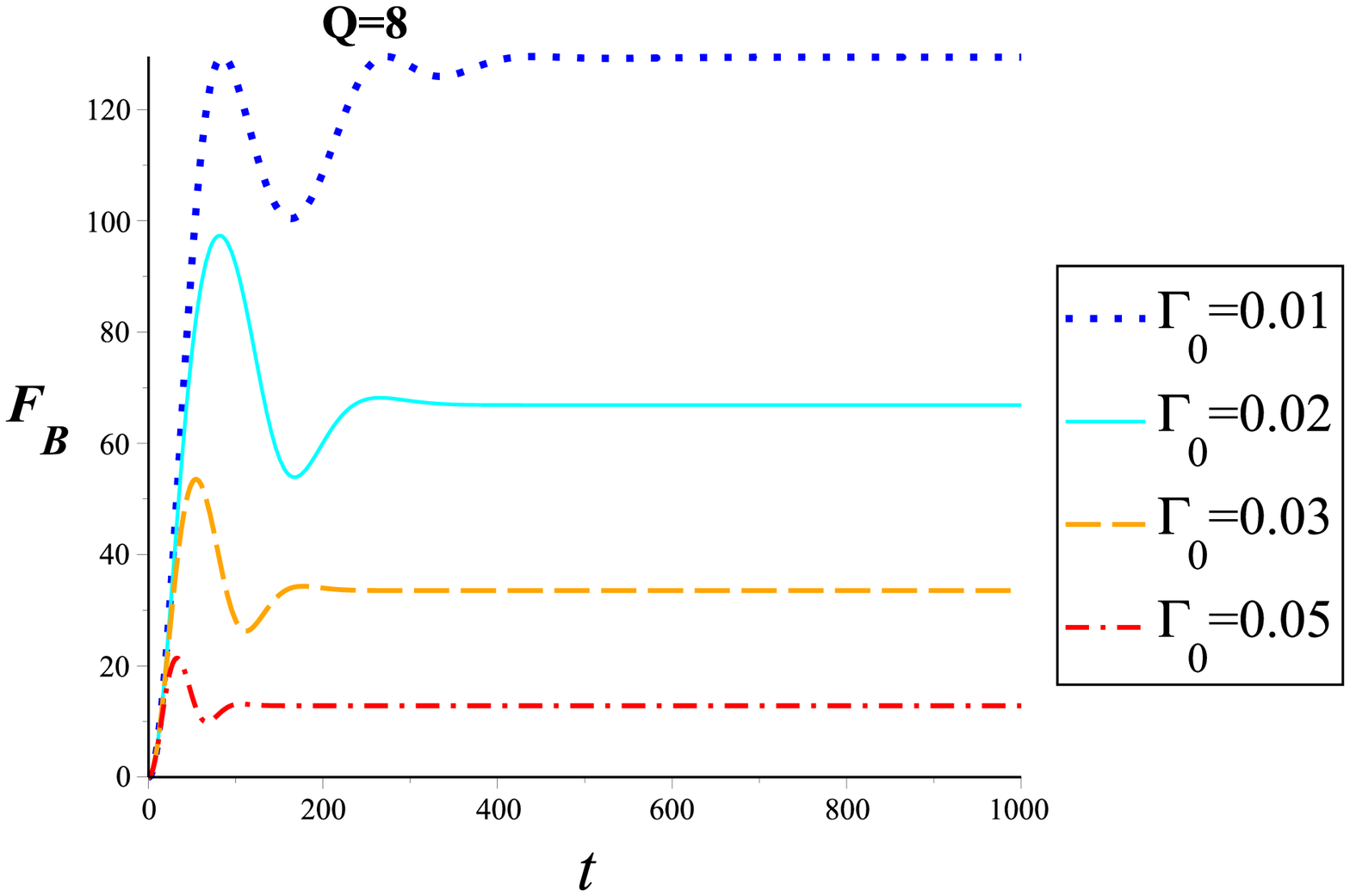}\label{QFIgama2} }
                       \caption{The time variation of the QFI associated with the magnetometry for different values of cutoff $ \varGamma_{0} $ in (a) Markovian and (b) non-Markovian regimes. }
                       \label{QFI1}
                         \end{figure}
                              \begin{figure}[ht!]
               \subfigure[]{\includegraphics[width=7cm]{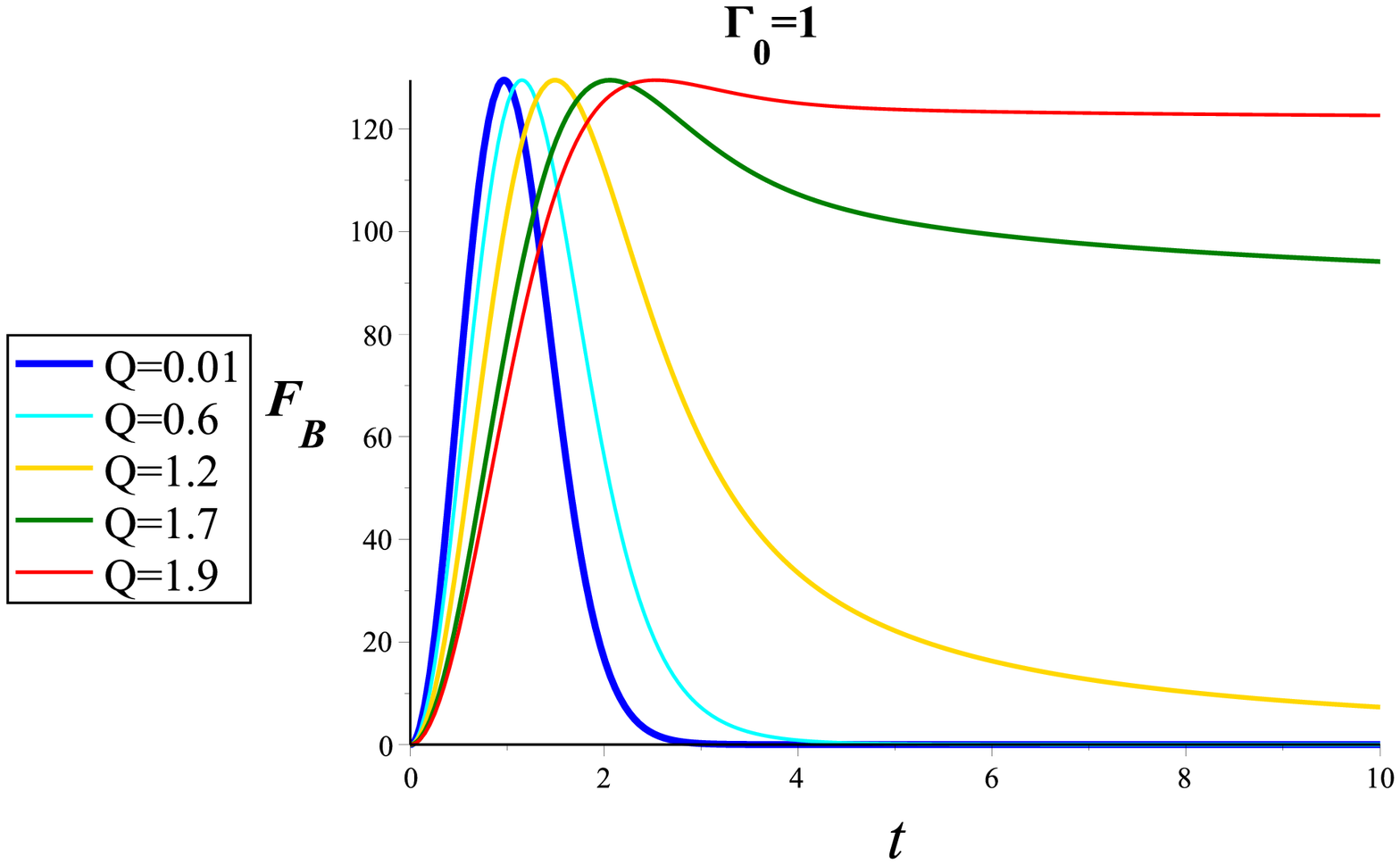}\label{QFIQ1} }
                \subfigure[]{\includegraphics[width=7cm]{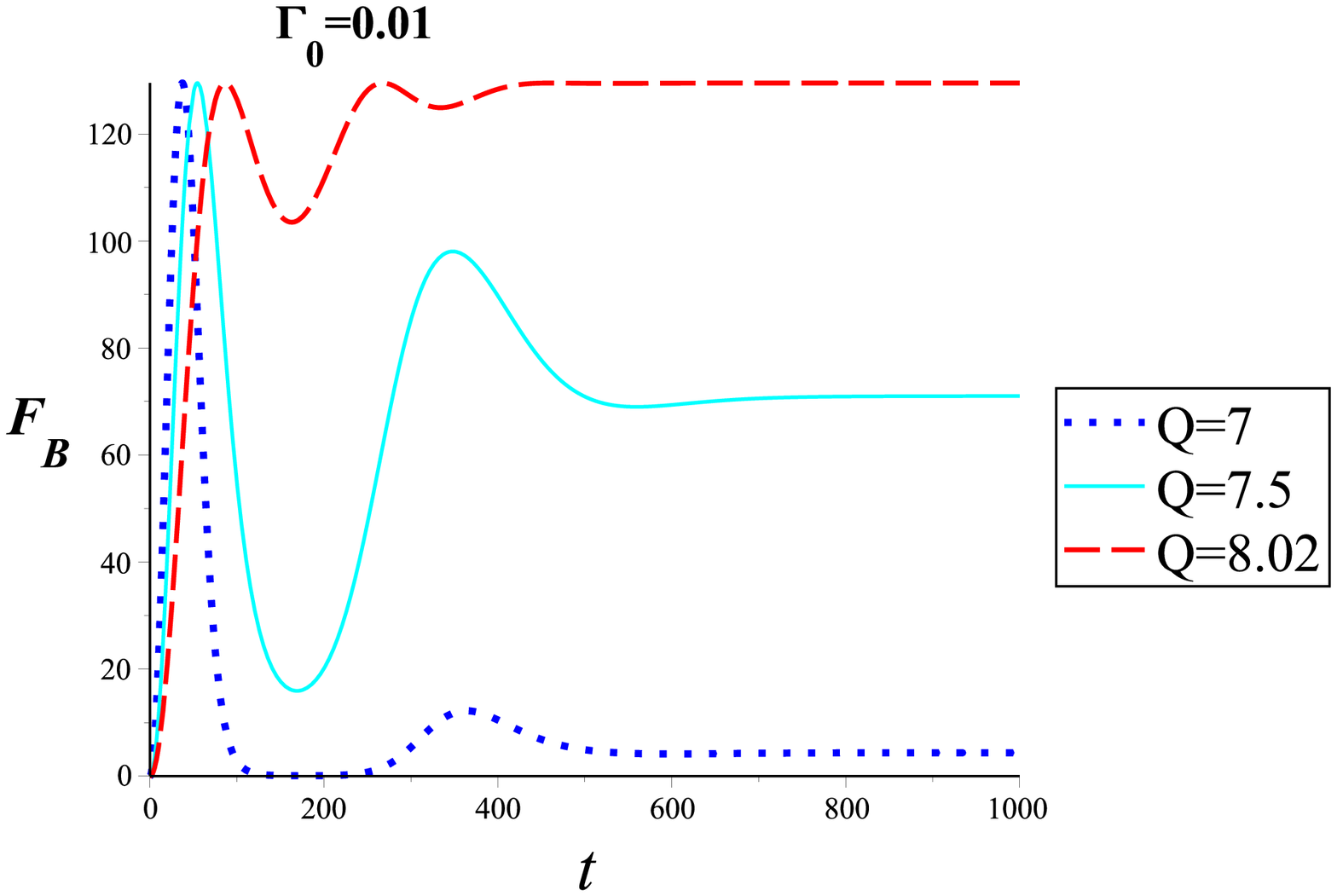}\label{QFIQ2} }
                              \caption{The time variation of the QFI associated with the magnetometry for different values of Ohmicity parameter $ Q $ in (a) Markovian and (b) non-Markovian regimes. }
                               \label{QFI2}
                                                  \end{figure}
      \par
       The  best  possible  precision  of  the parameter to be estimated  is  given  by  
        the Cram\'{e}r-Rao bound: Let us suppose that one performs $ N $ independent measurements in order to  achieve an
       \textit{unbiased estimator} $ \hat{\eta} $ \cite{Braunstein1996,RanganiAOP2}  for  parameter $ \eta $ (such that $ \text{Tr}[\rho_{\eta}\hat{\eta}]=\eta $). It can be shown that the
       variance of the estimator is lower bounded by 
       $ \langle (\hat{\eta}-\langle \hat{\eta}\rangle)  ^{2} \rangle \geq \dfrac{1}{NF_{\eta}}$  \cite{Braunstein1994}  where $ F_{\eta} $ denotes  the quantum Fisher information (QFI) \cite{Helstrom1976,Braunstein1994,Yao2014,RanganiOPTC,J1,J2,J3},  given by
       \begin{equation}\label{OrginalQFI}
       F_{\eta}=\sum_{i,j}\frac{2}{\omega_{i}+\omega_{j}}|\langle \phi_{i}|\frac{\partial \rho_{\eta}}{\partial \eta}|\phi_{j}\rangle|^{2},
       \end{equation}
       for the  mixed state with  spectral decomposition   $
       \rho_{\eta}=\sum_{i} \omega_{i}|\phi_{i}\rangle\langle\phi_{i}|
       $.
       
       \par                                       
           We consider a scheme in which the two-qubit system, prepared initially in maximally entangled state (\ref{initialtwoqubit}),
           is used to sense the intensity of the external magnetic field driving the topological qubits. The QFI associated with the magnetometry is given by:
           \begin{equation}\label{QFIASLI}
                           F_{B} =128\,{\frac {{\alpha}^{2}{B}^{2}|\beta|^{2}{I^{2}_{Q}}{~
                                      \text{e}^{-4\,{B}^{2} \left| \beta \right| I_{Q}}}}{1-{\alpha}^{4}}}.
                              \end{equation}
                             On one hand, the  information about the intensity of the magnetic field, encoded into the quantum state of the system, is able to flow into the environment. On the other hand, the system continously encodes the
                             information   because of the  interaction with the magnetic field. In Markovian regime, as illustrated in
           Fig. \ref{QFIgama1},  these two mechanisms compete with each other and in non-Markovian regime the quantum memory is added to this competition (see Fig. \ref{QFIgama2} ).  In both Markovian and non-Markovian regimes, the QFI initially increases  since the information is encoded into the quantum state  because of the interaction with the magnetic field. Then the decoherence effect dominates the dynamics of the topological qubits leading to loss of encoded information, although in non-Markovian evolution  the quantum memory may stop this process and help the interaction mechanism to temporarily increase the QFI. Moreover, the important role of the cutoff in controlling the dynamics and  extracting the  information  from the Majorana system is clear from Fig. \ref{QFI1}. It shows that in Markovian evolution,  with increasing cutoff $ \varGamma_{0} $, the QFI degradation can be frustrated. In fact, after a while, the QFI can remain approximately constant over time, a
                   phenomenon known as \textit{QFI trapping}. Although freezing of the QFI would be satisfactory and lets us protect the encoded information, the QFI may be suppressed by imposing larger values of the cutoff. In non-Markovian regime the QFI tapping  always appears after a while and the  QFI suppression also occurs  when the cutoff increases.
                    
           Now we focus on effects of the Ohmicity parameter on the dynamics of the QFI.
          Figure \ref{QFIQ1}
           exhibits an important and interesting consequence of an increase in parameter $ Q $ in Markovian regime ($ Q<2 $), leading to retardation of the QFI loss during the time evolution and
           therefore enhance the estimation of the parameter at a later time. Moreover,
           when the Ohmicity parameter varies from 0 to 2,  the optimal value of estimation is not affected by this variation. Nevertheless, varying $ Q $ from $ 0 $ to $ 2 $, we see that the  QFI reaches its maximum value at a later time-point, and hence the instant at which the optimal estimation occurs is achieved later. On the other hand, in non-Markovian regime we find that when the Ohmicity parameter approaches the value $ Q_{\text{opt}} $ at which the non-Markovianity   is maximized, the QFI can be   frozen at its optimal value (compare Figs. \ref{QFIQ2} and \ref{f1}). Overall, the best estimation is obtained when $ Q\approx Q_{\text{opt}} $ and 
             when $ Q\gg Q_{\text{opt}}   $, the QFI approaches zero.

  \section{Summary and conclusions \label{conclusion}}
  \par To summarize, we investigated the quantum correlations and  non-Markovian dynamics of Majorana qubits coupled to Ohmic-like fermionic environments and used for magnetometry.  The fermionic
environment is the helical Luttinger liquids realized
 as interacting edge states of two-dimensional topological insulators.
The influence functional $ \alpha $, appeared in the evolved density matrix of the topological qubit (see Eq. (\ref{Reduced1qMajorqubit})), leads to imposition of  a cutoff for the linear spectrum of the edge states, playing 
 an important role in the dynamic of the system. 
 In fact,  this influence functional is the exclusive characteristic of topological qubits \cite{HaoNJP},
 in contrast with the one for non-topological qubits. Hence, the introduced cutoff, which can be applied as a control parameter for non-Markovian evolution of the system,  results from the
 non-local nature of the topological qubits and the peculiar algebra of the Majorana modes.
We  discovered the
 existence of a cutoff-dependent critical value of the
 Ohmicity parameter for the Markovian to non-Markovian
 crossover.
 It was illustrated that the memory effects leading to information backflow and recoherence appear only if
  the reservoir spectrum is super-Ohmic with $ Q > 2 $, however this critical value can be controlled by  the cutoff. These results are consistent with ones obtained in \cite{Haikka0101032013} in which the authors discussed the existence of a temperature-dependent critical value of the Ohmicity parameter for the onset of non-Markovianity in local qubits experiencing  dephasing with an Ohmic class
  spectrum. However, our results are different from ones presented in \cite{HaoNJP} in which the advent of non-Markovianity for super-Ohmic environment driving the topological qubit, has been claimed for $   Q>1  $ and the independence of  presented results from the choice of the  cutoff  
  has been emphasized.

  \par
  We also investigated the ESD, disruptive to quantum information processing due to
  the fast disappearance of entanglement. It was shown that
  the ESD  may occur in both Markovian and non-Markovian dynamics. In Markovian regime, if the  ESD occurs the entanglement cannot be restored over time. However,   in non-Markovian regime 
  we observed the  counterintuitive
  entanglement rebirth after its sudden death or disappearance of the ESD  by increasing the cutoff,
  demonstrating the high level of quantum control  realized by imposing the cutoff.
   The entanglement sudden death or
  rebirth have been also implemented experimentally within linear optics set-up with photonic qubits \cite{Almeida5792007,Xu1005022010} and between atomic ensembles \cite{Laurat1805042007}.
  In Ref. \cite{Lopez0805032008}, it has been shown that the ESD in a bipartite
  system independently coupled to two reservoirs is necessarily related to the entanglement
  sudden birth between the environments. In fact, the
  loss of entanglement is related to the birth of entanglement
  between the reservoirs and other partitions. 
  The above study can easily be extended to investigate the
  entanglement dynamics starting from different initial conditions and to take into account finite temperature effects.

  We then  showed that the various quantum correlations and
 quantum coherence may be approximately frozen  for all time by imposing a large cutoff and hence the time-invariant quantum resources may be  realized with good approximation. The phenomenon of  freezing of quantum
   discord has not remained as a purely theoretical construct. It has been  several
   experiments demonstrating these peculiar effects with different physical systems such as optical experimental setup \cite{Xu72010} and  room temperature nuclear magnetic resonance
   setup \cite{Auccaise1404032011}.

  Finally, we investigated the effects of the cutoff and   Ohmicity parameter on the dynamics of the QFI. It was found that imposing large values of the cutoff is not satisfactory because of  the QFI suppression. On the other hand, for different values of  $ Q $, we illustrated that
   the QFI associated with the magnetometry initially increases.
            In \cite{X.-M.0421032010}, it was proposed that a positive QFI flow at time
            t, i.e., $\dfrac{dF}{dt}>0  $, implies that the QFI flows back into the system from the
            environment, generating a non-Markovian dynamics. However, in our estimation, the positivity of $\dfrac{dF_{B}}{dt}  $  does not necessarily detect the non-Markovianity \cite{RanganiJMO}, because we have focused on estimating one of the environment parameters.
              In fact, when the probes
             interact with the environment, the information about
             the intensity of the magnetic field is encoded into the quantum state of
             the probes, and hence the QFI increases. According to theory of quantum metrology, an increase
                          in the QFI indicates that the optimal precision of estimation
                          is enhanced. However,
                        simultaneously the interaction with the environment leads to flow of encoded information from the system to the environment. 
                        When this decoherence effect overcome the encoding process,
                         its destructive influence appears and we see that the QFI 
                        monotonously decays with time, resulting in the quantum magnetometry becomes more
                        inaccurate.

\section*{Acknowledgements}

 The authors are grateful to the anonymous referee for her/his comments and suggestions. H.R.J. would like to appreciate Rosario Lo Franco for useful discussions and correspondence. H.R.J. also wishes
 to acknowledge the financial support of the MSRT of Iran and
 Jahrom University.

\appendix

\section{Density matrix elements for two Majorana qubits independently interacting with the fermionic
environments }\label{DMB}
In the standard  basis
           $ \{|00\rangle \equiv |1\rangle ,|01\rangle \equiv |2\rangle,|10\rangle \equiv |3\rangle,|11\rangle\equiv |4\rangle\} $, using the approach introduced in \cite{Bellomoprl} and considering (\ref{InitialMajorqubit}), we find that the diagonal and  non-diagonal elements of the density matrix for the two-qubit system, respectively, are given by:

\begin{align}\label{MasterELEMENT1}
           \rho_{1,1}(t)=\frac{1}{4}\big(1+\alpha^{2}(t)\big)^{2}\rho_{1,1}(0)+\frac{1}{4}\big(1-\alpha^{2}(t)\big)^{2}\rho_{44}(0)+\frac{1}{4}\big(1-\alpha^{4}(t)\big)\big[\rho_{2,2}(0)+\rho_{3,3}(0)\big],
           \nonumber \\
           \rho_{2,2}(t)=\frac{1}{4}\big(1+\alpha^{2}(t)\big)^{2}\rho_{2,2}(0)+\frac{1}{4}\big(1-\alpha^{2}(t)\big)^{2}\rho_{3,3}(0)+\frac{1}{4}\big(1-\alpha^{4}(t)\big)\big[\rho_{1,1}(0)+\rho_{4,4}(0)\big],
           \\
          \rho_{3,3}(t)=\frac{1}{4}\big(1+\alpha^{2}(t)\big)^{2}\rho_{3,3}(0)+\frac{1}{4}\big(1-\alpha^{2}(t)\big)^{2}\rho_{2,2}(0)+\frac{1}{4}\big(1-\alpha^{4}(t)\big)\big[\rho_{1,1}(0)+\rho_{4,4}(0)\big]\nonumber, 
          \\
          \rho_{4,4}(t)=\frac{1}{4}\big(1+\alpha^{2}(t)\big)^{2}\rho_{4,4}(0)+\frac{1}{4}\big(1-\alpha^{2}(t)\big)^{2}\rho_{1,1}(0)+\frac{1}{4}\big(1-\alpha^{4}(t)\big)\big[\rho_{2,2}(0)+\rho_{3,3}(0)\big]\nonumber,
           \end{align}
           and
           
  \begin{align}\label{MasterELEMENT2}
         \rho_{1,2}(t)=\frac{1}{2}\alpha(t)\big(1+\alpha^{2}(t)\big)\rho_{1,2}(0)+\frac{1}{2}\alpha(t)\big(1-\alpha^{2}(t)\big)\rho_{3,4}(0),
            \nonumber \\
           \rho_{1,3}(t)=\frac{1}{2}\alpha(t)\big(1+\alpha^{2}(t)\big)\rho_{1,3}(0)+\frac{1}{2}\alpha(t)\big(1-\alpha^{2}(t)\big)\rho_{2,4}(0),
           \\\nonumber
              \rho_{2,4}(t)=\frac{1}{2}\alpha(t)\big(1+\alpha^{2}(t)\big)\rho_{2,4}(0)+\frac{1}{2}\alpha(t)\big(1-\alpha^{2}(t)\big)\rho_{1,3}(0), 
           \\
       \rho_{3,4}(t)=\frac{1}{2}\alpha(t)\big(1+\alpha^{2}(t)\big)\rho_{3,4}(0)+\frac{1}{2}\alpha(t)\big(1-\alpha^{2}(t)\big)\rho_{1,2}(0).\nonumber
        \end{align}
        Moreover,
         \begin{equation}
              \rho_{1,4}(t)=\alpha^{2}(t)\rho_{1,4}(0),~~~~\rho_{2,3}(t)=\alpha^{2}(t)\rho_{2,3}(0).
  \end{equation}
\section{Analytical expressions for QD and LQU}\label{QDLQU}
  \par
  Computation of QD for general states is not an easy task, however for a two-qubit system with $ X $-type structure states, the analytical expression of QD is available \cite{Wang C-Z}
  \begin{equation}
  QD(\rho_{AB})=\text{min}\left(Q_{1},Q_{2} \right), 
  \end{equation}
  where
  \begin{eqnarray}
  \nonumber Q_{j}=H\left(\rho_{11} +\rho_{33}  \right)+\sum_{i=1}^{4}\lambda_{i}\text{log}_{2}\lambda_{i}+D_{j},~~~\left(j=1,2\right),~~~~~~~~\\
  D_{1}=H\left(\frac{1+\sqrt{\left[1-2\left( \rho_{33} +\rho_{44}\right)  \right] ^{2}+4\left(|\rho_{14}|+|\rho_{23}| \right) ^{2}}}{2} \right), \\\nonumber
  D_{2}=-\sum_{i}\rho_{ii}\text{log}_{2}\rho_{ii}-H\left(\rho_{11} +\rho_{33}  \right),~~~~~~~~~~~~~~~~~~~~~~~~~~~~~~~~~\\\nonumber
  \nonumber H\left(x \right)=-x\text{log}_{2}x-\left(1-x \right)\text{log}_{2}\left(1-x \right),~~~~~~~~~~~~~~~~~~~~~~~~~~~~~~
  \end{eqnarray}
  in which the eigenvalues of the bipartite density matrix $ \rho_{AB} $ are given by
  \begin{eqnarray}
  \nonumber\lambda_{1,2}=\frac{1}{2}\left[\left(\rho_{11} +\rho_{44}  \right) \pm\sqrt{\left(\rho_{11} -\rho_{44} \right)^{2} +4|\rho_{14}|^{2}} \right] ,\\
  \lambda_{3,4}=\frac{1}{2}\left[\left(\rho_{22} +\rho_{33}  \right) \pm\sqrt{\left(\rho_{22} -\rho_{33} \right)^{2} +4|\rho_{23}|^{2}} \right].
  \end{eqnarray}
\par  
          For $\rho(t) $ whose elements are given in Eq. (\ref{MasterELEMENT}), the QD can be obtained as
          \begin{equation}\label{QDiscord}
             QD=\text{min}(Q_{1},Q_{2}),   
         \end{equation}
          where 
           \begin{equation}\label{Q1}
                        Q_{1}={\frac {\ln    \left( {\alpha}^{4}-1 \right)  -
                        {\alpha}^{4}\ln  \left( 1-{\alpha}^{2} \right) +{\alpha}^{2}
                         \ln  \left(1 -{\alpha}^{4} \right) + {\alpha}^{2}\left( {\alpha}^{2}-2
                         \right) \ln   \left( {\alpha}^{2}-1 \right) 
                          }{2~\ln  \left( 2 \right) }},
                    \end{equation}
                    and 
           \begin{equation}\label{Q2}
                                   Q_{2}={\frac {\ln    \left( {\alpha}^{4}-1 \right)  -
                                    \left( {\alpha}^{4}+1 \right) \ln  \left( {\alpha}^{4}+1 \right) +
                                    \left( {\alpha}^{2}-2 \right) {\alpha}^{2}\ln  \left( {\alpha
                                   }^{2}-1 \right)  + \left( {\alpha}^{2}+2 \right) {
                                   \alpha}^{2}\ln  \left( {\alpha}^{2}+1 \right) }{2~\ln  \left( 2 \right) 
                                   }}.
                               \end{equation}

Moreover, considering the definition of LQU in Eq. (\ref{LQU}), we should note that for a qubit-qudit system,  the quantification of non-classical correlations is not affected by  the choice
of the spectrum $ \varLambda $, therefore 
one can drop the $ \varLambda$ superscript. In addition, for qubit-qudit systems, it is possible to compute the LQU via the following relation \cite{Lecture}: 
          \begin{equation}
LQU_{A}=1-\lambda_{max}\big(W_{AB}\big),
          \end{equation}
in which $\lambda_{max}\big(W_{AB}\big)  $ denotes  the maximum eigenvalue of the $ 3\times 3 $ symmetric matrix $ W $ with
elements given by:
          \begin{equation}
(W_{AB})_{ij}=\text{Tr}\big[\sqrt{\rho_{AB}}(\sigma_{iA}\otimes \mathcal{I}_{B}) \sqrt{\rho_{AB}}(\sigma_{jA}\otimes \mathcal{I}_{B})\big ],
          \end{equation}
 in which the Pauli matrices are labelled by $ i, j $.
Using the above instruction, one can obtain the LQU as follows:

\[
    LQU= 
\begin{cases}
    1-\sqrt{1-\alpha^{4}}& \text{if }~ 2+{\alpha}^{4}\cos \left( 2\,\vartheta \right) \leq {\alpha}^{4}+2\
    \sqrt {1-{\alpha}^{4}}
    \\
    \alpha^{4}\text{sin}^{2}(\vartheta)             & \text{otherwise}.
\end{cases}
\]

\pagebreak

\end{document}